\documentclass[paper,letterpaper]{JHEP3} 
\usepackage{amssymb} 
\usepackage{epsfig}

\newcommand{\be}{\begin{equation}} 
\newcommand{\ee}{\end{equation}} 
\newcommand{\bea}{\begin{eqnarray}} 
\newcommand{\eea}{\end{eqnarray}} 
\newcommand{\ie}{{\it i.e.}} 
\newcommand{\eg}{{\it e.g.}} 
\newcommand{\Z}{\mathbb{Z}} 
\newcommand{\C}{\mathbb{C}} 
\newcommand{\CP}{{\C\mathbb{P}}} 
\newcommand{\U}{\mathop{\rm U}} 
\newcommand{\SU}{\mathop{\rm SU}} 
\newcommand{\SO}{\mathop{\rm SO}} 
\newcommand{\SL}{\mathop{\rm SL}} 
\newcommand{\Sp}{\mathop{\rm Sp}} 
\newcommand{\bo}{\hbox{1\kern-.23em{\rm l}}}
\newcommand{\til}{\widetilde}
\newcommand{\et}{{\til e}} 
\newcommand{\wt}{{\til w}} 
\newcommand{\xt}{{\til x}} 
\newcommand{\yt}{{\til y}} 
\newcommand{\zt}{{\til z}} 
 
\title{The M theory lift of two O6$^-$ planes and four D6 branes}

\author{Philip C. Argyres$^1$\thanks{On leave from Newman Laboratory, 
		Cornell University, Ithaca NY 14853}\ , 
	R. Maimon$^2$ and Sophie Pelland$^2$\\ 
        ${}^1$Department of Physics, University of Cincinnati, 
		Cincinnati OH 45221-0011\\
	${}^2$Newman Laboratory, Cornell University, Ithaca NY 14853} 

\abstract{We solve for the effective actions on the Coulomb branches
of a class of $N=2$ supersymmetric theories by finding the complex
structure of an M5 brane in an appropriate background hyperkahler
geometry corresponding to the lift of two O6$^-$ orientifolds and four
D6 branes to M theory.  The resulting Seiberg-Witten curves are of
finite order, unlike other solutions proposed in the literature.  The
simplest theories in this class are the scale invariant $\Sp(k)$
theory with one antisymmetric and four fundamental hypermultiplets and
the $\SU(k)$ theory with two antisymmetric and four fundamental
hypermultiplets.  Infinite classes of related theories are obtained by
adding extra $\SU(k)$ factors with bifundamental matter and by turning
on masses to flow down to various asymptotically free theories.  The
$N=4$ supersymmetric $\SU(k)$ theory can be embedded in these
asymptotically free theories, allowing a derivation of a subgroup of
its S duality group as an exact equivalence of quantum field
theories.}
 
\preprint{UCTP-102-02, CLNS-02/1783} 

\begin{document}

\section{Introduction and summary}

A large class of supersymmetric gauge field theories can be realized
as open string excitations confined to branes in string theory.
Following \cite{w9703} the low energy effective actions on the Coulomb
branches of $N=2$ supersymmetric four dimensional theories with
product $\SU$ gauge groups and fundamental matter were constructed by
using IIA/M theory duality to realize the Seiberg-Witten curve
\cite{sw9407,sw9408} as the complex structure of an M5 brane.  This
construction was generalized in \cite{ll9708} to include a matter
hypermultiplet in a symmetric or an antisymmetric representation of
the gauge group by adding an orientifold O6 plane to the configuration
of D4, NS5, and D6 branes \cite{w9703,ll9708,u9803} (see also
\cite{ty9803}).  Theories with two such matter representations can
also be realized as a IIA brane configuration by adding a second O6
plane \cite{u9803}; however the resulting lift to an M5 brane curve
was thought to be more difficult in this case.  Furthermore, an
alternative approach based on ``reverse engineering'' the curve from
computations of one-instanton corrections to the Coulomb branch
effective coupling led to an infinite order Seiberg-Witten curve
\cite{enrs9904,elns9912,elnrs0006}.

In this paper we carry through the lift of a IIA string configuration
of two O6$^-$ planes and four D6 branes to an M theory gravitational
background.  We then show how to write curves describing appropriate
holomorphic embeddings of M5 branes in this background geometry so as
to preserve 8 global supersymmetries, corresponding to $N=2$
supersymmetric theories in four dimensions.  

The IIA configurations we are interested in all involve various D4 and
NS5 branes (whose geometry is detailed in section~2) in the presence
of a parallel set of two O6$^-$ orientifold planes and four D6 branes.
The heavy branes---the two O6$^-$ planes and the four D6
branes---form a background in which many NS5 and D4 brane
configurations can be placed.  We identify the various $N=2$
supersymmetric four dimensional weakly coupled gauge theory limits
that are obtained in this way.  They include the $\Sp(k)$ theory with
four fundamental hypermultiplets and one traceless-antisymmetric
matter hypermultiplet with arbitrary masses, the $\SU(k)$ theory with
four fundamental and two antisymmetric hypermultiplets, the $\Sp(k)
\times \Sp(k)$ theory with two fundamentals in each factor and one
bifundamental hypermultiplet, as well as infinite classes of related
theories that are obtained by adding extra $\SU(k)$ factors with
bifundamental matter and by turning on masses to flow down to various
asymptotically free theories.

The transverse space to the heavy branes---the two O6$^-$ planes
and the four D6 branes---lifts in M theory to a smooth hyperkahler
four-manifold $Q_0$, which we will refer to as the background surface.
In section~3 we show that one of its complex structures can be
described as a surface in the 3 complex dimensional space
$\CP^2_{(1,1,2)}\times\C$ coordinatized by $(w,x,y) \in
\CP^2_{(1,1,2)}$ with
\be
(\lambda w, \lambda x, \lambda^2 y)  
\simeq (w,x,y), \qquad \lambda \in \C^*,
\ee
and $z \in \C$.  The equation of the surface $Q_0$ is then given by
\be\label{introQ} 
y^2 = z \prod_{i=1}^4 (x-e_iw) + \sum_{j=1}^4 \mu_j^2 w  
\prod_{k\neq j} [(x-e_kw)(e_j-e_k)] . 
\ee
Only one combination of the $e_i$ is an invariant of the complex 
structure of $Q_0$, and has the interpretation as a complex gauge 
coupling of the associated scale invariant $N=2$ theories.  All four 
of the $\mu_j$ are complex structure invariants, and have the 
interpretation of linear combinations of the bare masses of certain 
hypermultiplet matter fields in the $N=2$ theories.  Indeed the 
$\mu_j$ are the residues of the poles of a meromorphic Seiberg-Witten 
one form on $Q_0$, given by 
\be 
\lambda = { y(wdx-xdw) \over \prod_i(x-e_i w) } . 
\ee 
Thus the $\mu_j$ and the $e_i$ coordinatize a total of five complex 
deformations of $Q_0$.

There is another deformation of this background geometry, associated
with a complex parameter $M$ which is important in the $N=2$ gauge
theories.  It is closely related to the ``shift'' of the elliptic
models introduced in \cite{w9703}.  In Section~3.4 we describe how to
implement this shift of $Q_0$, which we denote $Q_M$, in an indirect
way by specifying a submanifold of $Q_0$ to be excised, and the
modified boundary conditions that holomorphic functions on $Q_M$ must
satisfy at this submanifold.  In terms of the four-dimensional
physics, the parameter $M$ of this shift corresponds to the mass of an
antisymmetric hypermultiplet.

The particular arrangements of NS5 and D4 branes in the IIA string
theory described in section~2 are lifted to a single M5 brane in
M theory.  The construction of the complex structure of the curves
describing the embedding of the M5 brane in $Q_M$ is given in
Section~4.  In the case of no shift, these complex curves in $Q_0$ can
be written in the general form of sums of polynomials in $z$ times
meromorphic functions in $\CP^2_{(1,1,2)}$.  The order of the
polynomials in $z$ corresponds to the number of D4 branes (the rank of
the gauge group factors), while the positions of the poles in
$\CP^2_{(1,1,2)}$ encode the positions of the NS5 branes
(corresponding to the relative strengths of the gauge couplings of the
gauge group factors).  The complex deformation parameters appearing in
these curves have the interpretation of Coulomb branch vevs and
bifundamental masses in the field theory.  In the case where the shift
parameter is turned on, the curves lie in $Q_M$ and are more
complicated, though explicit prescriptions for their construction are
given in Section~4.3.  In particular, all the relevant formulas are
collected in equation (\ref{QMcur})---(\ref{final}).  The resulting
curves are all finite genus Riemann surfaces.

As a check on our construction, and for later use, in section~5 we
match our curves at weak coupling to perturbative results and to known
curves of other theories in certain decoupling limits.  This
determines explicit mappings between the various complex parameters
appearing in the background and M5 brane curves and the physical
couplings, masses, and vevs.  In this section we also derive some
curves for asymptotically free theories which can be found upon
decoupling fundamental hypermultiplets by sending their masses to
infinity.  In the IIA picture this corresponds to sending the D6
branes to infinity, leaving only the two O6$^-$ planes as the
background geometry.

In section~6 we show that by going to an appropriate submanifold of
the Coulomb branch of certain of these asymptotically free theories
(and by also turning on appropriate hypermultiplet vevs) we can embed
the $N=4$ supersymmetric $\SU(n)$ superYang-Mills theory in $N=2$
supersymmetric theories.  Although the enhanced $N=4$ supersymmetry is
an accidental symmetry at long wavelengths, this embedding can be used
to derive a subgroup of the S duality group of the $N=4$ theory,
following the general arguments of \cite{a9706}.  Assuming the
validity of the curves derived in section~6 for the low energy
effective action of the $N=2$ theories on the Coulomb branch, we
derive the complex structure of the image of the coupling space of the
embedded $N=4$ theory.  This, together with a global discrete symmetry
of the asymptotically free theory are enough to imply that the $N=4$
theory has a subgroup of the $N=4$ S duality group.  Furthermore, by
taking a scaling limit towards the singular submanifold of the Coulomb
branch of the asymptotically free theory, we can show, to any given
precision, that all the correlators of the $N=4$ (and not just the low
energy effective action or supersymmetric states) are duality
invariant under this subgroup.
 
Finally, it would be interesting to compare the predictions of the
curves found in this paper with those of the alternative infinite
order ones proposed in \cite{enrs9904,elns9912,elnrs0006}.  It is
possible that the two sets of curves are equivalent in the sense that
there is an infinite group of identifications on the
infinite-dimensional Jacobian torus of the infinite order curves under
which they becomes equivalent to the Jacobians of our curves; however,
it seems difficult to find such a ``folding'' in practice.  One test
of the equivalence of the two sets of curves is to compare their
predictions for the multi-instanton contributions to the prepotential.
Such a check would have to take into account the full set of allowed
non-perturbative redefinitions of the parameters of the two curves as
in \cite{ap9911}.

As this paper was being written up, the paper \cite{kn0203} appeared
which also discusses curves describing branes in the presence of 
two O6$^-$ planes and four D6 branes in the special case with
the deformation parameters (fundamental masses) $\mu_i=0$.

\section{IIA string construction}

We begin by describing a set of IIA string constructions whose weak
string coupling and low energy limits describe the set of
four-dimensional scale invariant $N=2$ supersymmetric gauge theories
we are interested in.

Let $x_0$---$x_9$ be the coordinates on the ten-dimensional spacetime
of type IIA string theory with $x_6$ compactified on a circle.  The
number and type of IIA string theory objects appearing in the
configuration, as well as the dimensions
along which they extend, are as follows:
\begin{itemize}
\item 2 O6$^-$ orientifolds along $x_0$---$x_3$, $x_7$---$x_9$, 
\item 4 D6 branes along $x_0$---$x_3$, $x_7$---$x_9$, 
\item $q$ NS5 branes along $x_0$---$x_5$ and intersecting an O6$^-$, 
\item $r$ NS5 branes along $x_0$---$x_5$, not intersecting any O6$^-$,
\item $k$ D4 branes along $x_0$---$x_3$, $x_6$. 
\end{itemize}
As an example, figure~1 shows an $x_4$---$x_6$ cross-section of a
fundamental domain of a configuration with $q=2$, $r=1$, and $k=2$ after
modding out by the orbifold identifications associated with the
orientifold planes.  The two ends of the domain in $x_6$ are identified,
and correspond to the location of one of the $q=2$ NS5 branes, with the
other midway between the two ends along $x_6$. The intersections of the
$q=2$ NS5 branes with the diagonal boundary of the fundamental domain
are two $\Z_2$ orbifold points corresponding to the positions of the O6
planes.  D4 branes suspended between each pair of NS5 branes are shown.
As they cross the diagonal boundary of the domain, they emerge on the
other side of boundary (where the middle NS5 naturally partitions the
domain into two sides).

\FIGURE{
\epsfig{file=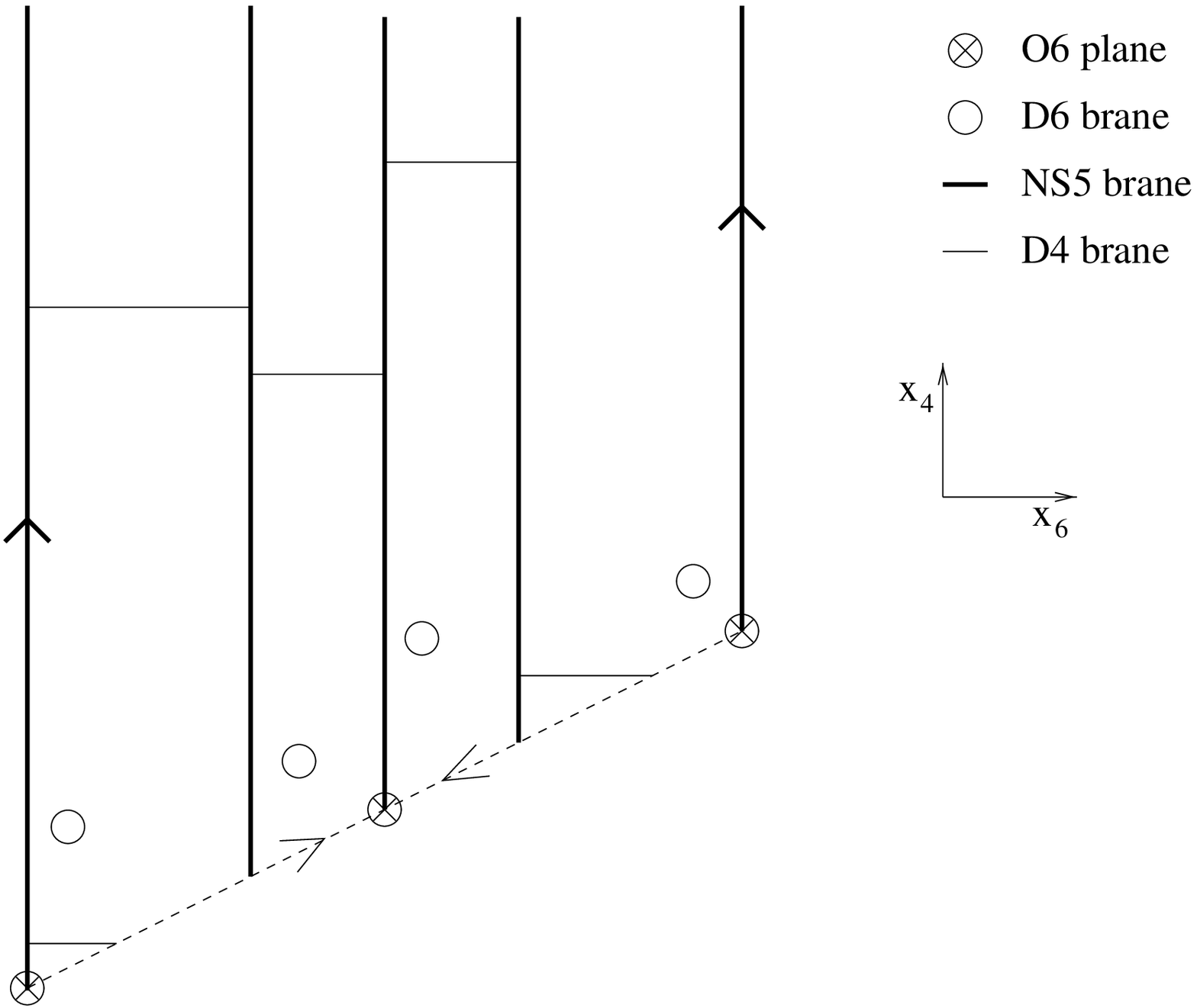,width=20em}	
\caption{A fundamental domain in the 
$x_4$---$x_6$ plane of an arrangement of two O6$^-$, three NS5 (two of 
which intersect the O6$^-$'s), two D4 (each split into two segments), and 
four D6 branes.}}

More explicitly, the O6$^-$ planes have the same $x_4$-$x_5$
coordinates, but they are separated along a diameter of $x_6$.  The D6
branes are parallel to the O6$^-$ planes.  Define $v \equiv x_4+ix_5$,
and let $2L$ be the circumference of the $x_6$ circle.  If we put one
O6$^-$ plane at $(v,x_6)=(0,0)$ and the other at $(L,M/2)$, then the
$x_6$-$v$ space is orbifolded by the identifications:
\be\label{orbi1} 
x_6 \simeq -x_6\quad\mbox{and}\quad v \simeq -v, 
\ee 
and 
\be\label{orbi2} 
x_6 \simeq x_6+2L\quad\mbox{and}\quad v \simeq v+M. 
\ee 
We will refer to $M$ as the ``shift'' in $v$ in the remainder of the 
paper. 

The NS5 branes have common $x_7$---$x_9$ coordinates but different
$x_6$ coordinates.  We count the $r+q$ NS5 branes as follows.  If it
does not intersect an O6$^-$ plane, then we count as contributing one
to $r$ a NS5 brane together with its image under (\ref{orbi1}); thus
in the orbifolded background $r$ counts the asymptotic NS5 brane
charge.  If it intersects an O6$^-$ plane, then it has no image 5
brane, and we count it as contributing one to $q$, even though it
contributes only $1/2$ to the asymptotic 5 brane charge.  We consider
only $q=0$, 1 or 2 where at most one of the NS5 branes coincide with a
given O6$^-$ plane.  The interesting cases with two or more NS5 branes
coinciding with a given O6$^-$ plane will be discussed elsewhere.

For the scale invariant models, the D4 branes wrap the $x_6$ circle.
Since they have the same $x_7$---$x_9$ coordinates as the NS5 branes,
they can split into a set of segments suspended between pairs of NS5
branes.  It will prove convenient to let $k$ count the number of D4
branes wrapping the $x_6$ circle {\em before} the orbifolding
(\ref{orbi1}), \ie\ a D4 brane and its image under (\ref{orbi1})
contributes {\em two} units to $k$.  The reason for this is that two
different segments of one D4 brane can end up, upon orbifolding, being
suspended between the same pair of NS5 branes.

This is illustrated in figure~2(a), where the snocone geometry of
figure~1 is made explicit in the simple case where $M=0$ (the
non-shifted case).  A snocone is a semi-infinite cylinder-like object
with a closed end, flat except at the two $\Z_2$ orbifold points.  Note,
for example, that there are two D4 branes suspended between the middle
U-shaped 5 brane and the one on the right side (which ends on an
O6 plane).  Figure~2(b) attempts to show the shifted case which is
equivalent to introducing a relative tilt between the directions in
which the NS5 and the snocone surfaces extend.
 
\FIGURE{
\epsfig{file=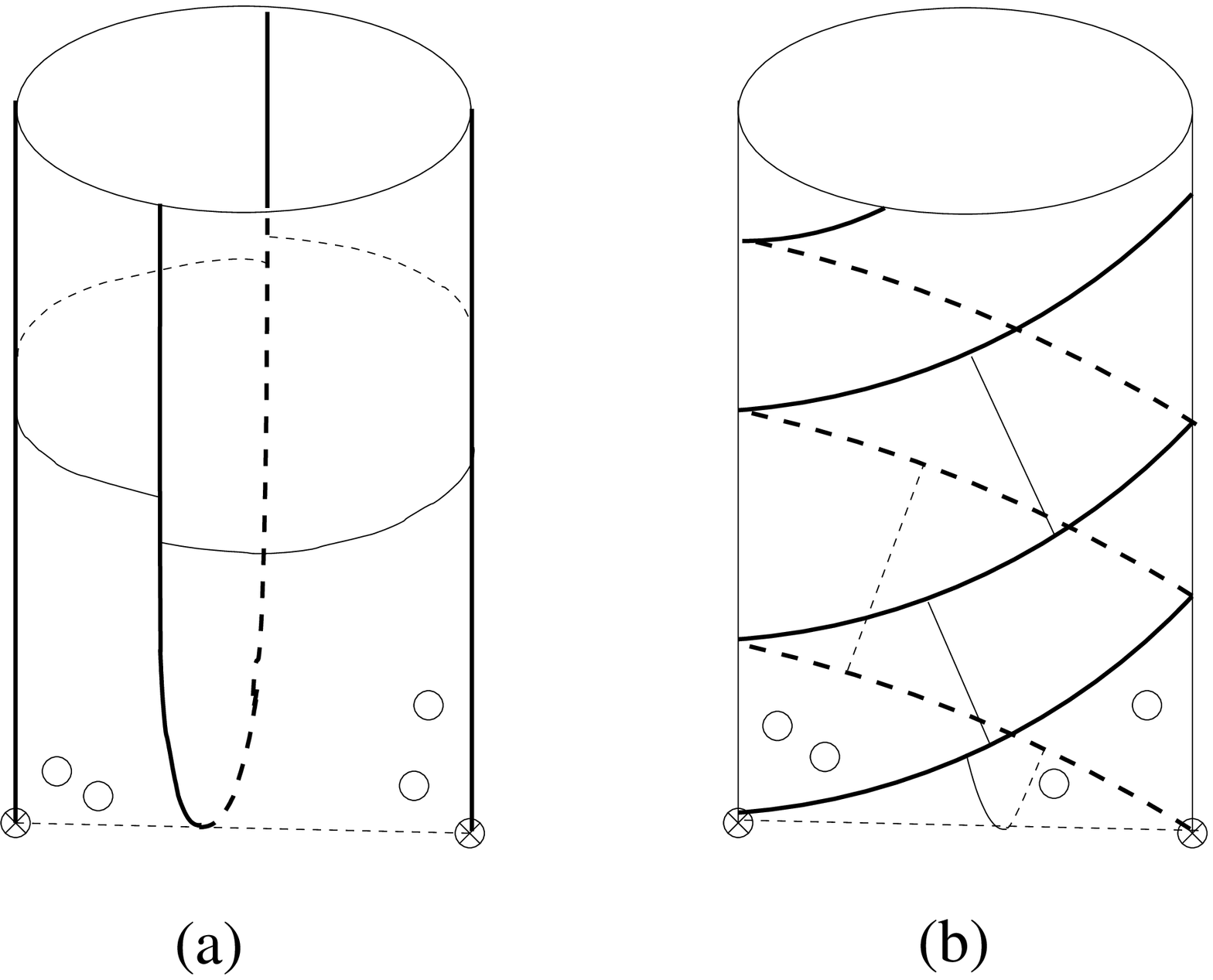,width=20em}
\caption{The topology of the  
region shown in figure~1 in the cases of (a) $q=2$, $r=1$, $k=2$,
and $M=0$, and (b) $q=2$, $r=0$, $k=4$, and $M\neq0$.}}

The low energy excitations of open strings confined to the D4 branes
of the above configurations give rise to the Coulomb branch of low
energy theories in four dimensions, since the brane configurations are
translationally invariant along $x_0$---$x_3$.  Each stack of $k$ D4
branes suspended between two NS5 branes results in an $\SU(k)$ gauge
factor \cite{w9510}, unless there is also an O6$^-$ plane between the
two NS5 branes, in which case the gauge factor is $\Sp(k)$
\cite{ll9708}.  (We use the notation where $\Sp(k)$ has rank $k/2$.)
Thus with $r$ NS5 branes (not intersecting an O6 plane) there will be
$r+1$ gauge group factors; if there are no 5 branes intersecting an O6
plane ($q=0$) then the ``end'' group factors are $\Sp(k)$, but if a 5
brane does intersect an O6 plane then that end factor is changed to
$\SU(k)$.

Matter hypermultiplets enter in three ways.  Firstly, fundamental
hypermultiplets arise from D6 branes passing between NS5 branes, and
their excitations correspond to strings extending from the D4 brane
stack to the D6 brane.  The mass of the fundamental is therefore
proportional to the distance in the $x_4$-$x_5$ plane of the D6 brane
from the D4 stack.  The D6 branes are associated in pairs to the two
O6 planes in our construction, and should be thought of as
contributing a pair of fundamental hypermultiplets to each of the
gauge group factors (D4 brane stacks) nearest the O6 planes.
Secondly, antisymmetric matter hypermultiplets arise when an O6$^-$
brane intersects an NS5 brane, and correspond to strings extending
from the D4 brane stack to the O6$^-$ plane.  The mass of the
antisymmetric is therefore proportional to the distance in the
$x_4$-$x_5$ plane of the O6$^-$ plane from the D4 stack.  Finally,
neighboring gauge factors have a bifundamental hypermultiplet from
strings stretching between neighboring stacks of D4 branes, and their
masses are proportional to the distance in the $x_4$-$x_5$ plane
between the two neighboring D4 stacks.

Since the 6 brane charges cancel and each 5 brane has equal numbers of
4 branes extending to the left and right, the asymptotic 5 branes are
flat, implying that the corresponding gauge groups should all be scale
invariant.  In particular, the asymptotic separations along $x_6$
between the $r$ non-stuck 5 branes as well as the O6 planes correspond
to the dimensionless couplings of the various gauge group factors.
Upon lifting to M theory, the 5 brane separations along the extra
$x_{10}$ circle correspond to the theta angles of the gauge factors.

To summarize, the parameters of the IIA configurations are identified
with parameters and vevs of the four dimensional gauge theories as
follows \cite{w9703,ll9708,u9803}.  The coupling of each gauge group
factor is identified with the separation of the pair of neighboring
NS5 branes associated with it. The scalar components of the adjoint
multiplet correspond to the relative motions of pairs of D4 branes.
Meanwhile, the masses of the fundamental, antisymmetric and
bifundamental multiplets correspond respectively to the relative
motions of D4 branes and D6 branes, D4 branes and O6$^-$ planes and
mean positions of the D4 branes on either side of an NS5 brane. Thus,
in particular, the parameter $M$ can be interpreted as the difference
of the masses of the two antisymmetric hypermultiplets.

The resulting $N=2$ gauge theories can be classified by the number
($q$) of NS5 branes which intersect the two O6$^-$ planes.  We list
here the gauge group and hypermultiplet content for the $q=0,1,2$
theories.  We also count all the complex parameters describing
the Coulomb branch geometry of each theory for later use.
\begin{itemize} 
\item $q=0$: $\Sp(k)\times\SU(k)^{(r-1)}\times\Sp(k)$ with 
bifundamentals between each pair of neighboring gauge group factors 
and two fundamentals in each $\Sp(k)$ factor.  There are $r+1$
couplings, 4 fundamental masses, $r$ bifundamental masses, and
$2(k/2) + (r-1)(k-1)$ Coulomb branch vevs, for a total of $6+r+kr$
parameters.
\item $q=1$: $\Sp(k) \times \SU(k)^{(r-1)} \times \SU(k)$ with 
bifundamentals between neighboring pairs, two fundamentals in 
$\Sp(k)$ and two fundamentals and an antisymmetric in the last 
$\SU(k)$ factor.   There are $r+1$ couplings, 4 fundamental masses, 
$r$ bifundamental masses, 1 antisymmetric mass, and $(k/2) + r(k-1)$ 
Coulomb branch vevs, for a total of $6+r+kr+(k/2)$ parameters.
\item $q=2$: $\SU(k) \times \SU(k)^{(r-1)} \times \SU(k)$ with 
bifundamentals between neighboring pairs, and two fundamentals and an 
antisymmetric in the first and last $\SU(k)$ factors.  There are $r+1$ 
couplings, 4 fundamental masses, $r$ bifundamental masses, 2 
antisymmetric masses, and $(r+1)(k-1)$ Coulomb branch vevs, for a total 
of $6+r+kr+k$ parameters.
\end{itemize}
In the $q=0$ and $q=1$ cases above, $k$ is even, since there is no
$\Sp(k)$ theory with $k$ odd.  In the $q=2$ case, on the other hand, odd
$k$ is allowed.  In terms of the brane configuration, because of the 5
branes intersecting each O6 plane, having an odd number of D4 branes
(recall that $k$ counts both D4 branes and their orbifold images) does
not mean that one is stuck on the O6 plane, since it can move by
splitting along the stuck NS5 brane.  The curves we will derive in
section~4 incorporate these rules in a natural way. 

The above list degenerates for $r=0$ as follows \cite{u9803}:
\begin{itemize} 
\item $q=2$: $\SU(k)$ with 2 antisymmetric hypermultiplets and 
4 fundamentals, which has 1 coupling, 4 fundamental masses,
2 antisymmetric masses, and $k-1$ vevs for a total of $k+6$ parameters. 
\item $q=1$: $\Sp(k)$ with a traceless-antisymmetric and 4 fundamentals,
which has 1 coupling, 4 fundamental masses, 1 antisymmetric mass, and 
$(k/2)$ vevs for a total of $(k/2)+6$ parameters. 
\item $q=0$: also yields $\Sp(k)$ with a traceless-antisymmetric and 
4 fundamentals, but only for $M=0$ (no shift) which corresponds to 
the antisymmetric hypermultiplet being massless.  Thus this model has
a total of $(k/2)+5$ parameters.
\end{itemize}
The $q=r=0$ configuration only exists for zero shift because otherwise
the 4 branes would not close upon traversing the $x_6$ circle.  The
equivalence of the $q=r=0$ and $q=1$, $r=0$
configurations at zero shift reflects the fact that the stuck 5
branes can move off along the O6 plane when the D4 branes close upon
traversing the $x_6$ circle.  In terms of the four dimensional gauge
theory this reflects the existence of a mixed Coulomb-Higgs branch at
$M=0$ which includes the whole Coulomb branch as a subvariety.  Even
with $M\neq 0$, if the D4 branes line up on either side of one of the
stuck 5 branes in the $q=2$ $r=0$ theory, that 5 brane can also be
moved off, giving the $q=1$ $r=0$ theory.  In field theory terms,
giving a vev to one of the antisymmetrics in the $\SU(k)$ theory
Higgses it to $\Sp(k)$ with one antisymmetric.  These and many other
relations between the theories we are considering found by tuning
masses and vevs and taking decoupling limits are reflected in the
structure of the curves we derive below.

{}From the counting of couplings, masses, and vevs given above, we see
that the total number of parameters on the Coulomb branch is
$6+r+kr+(qk)/2$ (except for the degenerate $r=q=0$ model).  Also, all
these theories have $r+1$ couplings and 4 fundamental masses.  This
counting will provide a simple check of the curves found in section~4
below.

Finally, it will be important for a later argument to consider the
effect of moving a D6 brane along a cycle enclosing one of the $\Z_2$
orbifold points, which corresponds to taking $m_i \to e^{ 2\pi i} m_i$
where $m_i$ is the fundamental mass parameter associated with the
chosen D6 brane.  As the D6 brane crosses an NS5 brane, it pulls a D4
brane with a definite orientation behind it (the type IIA equivalent
of the Hanany-Witten effect \cite{hw9611}).  Moving the D6 brane
across the diagonal boundary in figure~1, we obtain a D6
brane on the other side with the D4 brane orientation reversed.
Finally, as the D6 traverses the NS5 brane a second time, a D4 brane
with opposite orientation to the first D4 brane is created, and
annihilates the first D4, leaving only a D6 brane on the other side of
the NS5.  The D6 can then recross the diagonal boundary and return to
its original position. Thus, only when the D6 circles the O6$^-$ twice
is the original configuration recovered.  We will return to this point
in section~4 to resolve an apparent sign ambiguity in our curves.

\section{M theory lift of two O6$^-$ planes}
 
The next step is to lift the IIA string constructions described above
to M theory \cite{w9703}, thereby obtaining the low energy effective
action on the Coulomb branch of the associated $N=2$ gauge theories.
All the IIA brane setups have the same arrangement of two O6$^-$
planes and four D6 branes which are transverse to the $x_4$, $x_5$,
and $x_6$ directions.  The lift of this transverse geometry to M
theory will give a hyperkahler four-manifold, $Q_M$, which we call the
background geometry or surface (since it is two complex dimensional).
In lifting to M theory, the D6 branes become multi-center Taub-NUT
manifolds, while the O6$^-$ planes become Atiyah-Hitchin manifolds
\cite{ah88,s9707,sw9607,s9606}.  We focus only on the complex
structure of this space, and do not determine its metric.
 
Meanwhile, all the NS5 and D4 branes lift to a single M5 brane, two
directions of which are embedded on a complex curve in the background
geometry $Q_M$.  Thus we adopt a convenient splitting of the M theory
lift into two steps: in this section we derive the hyperkhaler geometry
$Q_M$ associated with the O6$^-$ and the D6 branes, and then in the next
section we embed the M5 brane corresponding to our D4 and NS5 brane
configuration in $Q_M$.  We subdivide both steps of the lift into three
parts as follows.  First, we consider the case where a pair of D6 branes
is coincident with each O6$^-$ and $M$ is set to zero (no shift), so
that the M theory lift is simply an orbifold of the $x_6$-$x_{10}$
torus cross the $v$ plane (section~3.1).  Next, we allow the D6 branes
to move off the O6$^-$, introducing deformations of the orbifolded torus
corresponding to four fundamental hypermultiplet masses (section~3.2).
Finally, we discuss how to account for non-zero shift $M$ in the
background (section~3.4).  Along the way, we will derive the
Seiberg-Witten one-form needed to complete our description of the low
energy action, and relate it to parameters in the background geometry
(section~3.3). 
 
\subsection{$\C\times T^2/\Z_2$: Orbifolding the elliptic model} 
 
While the M theory lift of individual D6 branes and O6$^-$ planes
involves non-trivial geometries, the lift of an O6$^-$ plane coincident
with a pair of D6 branes is simply a $\Z_2$ orbifold known as a $D_2$
singular space \cite{sw9607}. It is therefore convenient to begin by
considering the $\Z_2$ orbifold of the $v$-plane cross the torus, and
then use physical requirements to constrain the deformations of the
orbifolded torus corresponding to moving the D6 branes off the O6$^-$
planes and to separating the O6$^-$ planes along $v$. 
 
In lifting to M theory, we grow a circular dimension $x_{10}$ of 
radius $R$.  Define the flat variable 
\be 
s \equiv (x_{10}+i x_6)/(2\pi R) . 
\ee 
Thus the orbifold identifications (\ref{orbi1}) and (\ref{orbi2}) are
lifted to 
\be\label{or3} 
s \simeq -s\quad\mbox{and}\quad v \simeq -v, 
\ee 
\be\label{or4} 
s \simeq s+1\quad\mbox{and}\quad v \simeq v, 
\ee 
and 
\be\label{or5} 
s \simeq s+\tau\quad\mbox{and}\quad v \simeq v+M, 
\ee 
where $\tau$ encodes the complex structure of the torus. 
In this and the next two subsections we will set the shift
parameter $M=0$.

Before orbifolding, the M theory background space $\til Q$ is just the
flat space $\til Q = \C \times T^2$ with coordinates $v\in \C$ and
$s\in T^2$.  We wish to orbifold this space under the $\Z_2$
identication $(v,s) \simeq (-v,-s)$, which can be done by rewriting
$\til Q$ in terms of orbifold-invariant coordinates, thus providing
good coordinates on $\til Q/\Z_2$.
 
We first introduce a single-valued coordinate description of $T^2$ by 
writing the compact torus as a complex curve in the weighted 
projective space $\CP^2_ {(1,1,2)}$ defined as the space of all 
complex $(w,x,\eta)$ minus the point $(0,0,0)$ and modulo the 
identification 
\be 
(\lambda w, \lambda x, \lambda^2 \eta) \simeq (w,x,\eta), 
\qquad \lambda\in \C^* . 
\ee 
Then we can write the torus as 
\be\label{torus2} 
\eta^2 = \prod_{i=1}^4 (x-e_i w) 
\ee 
where the numbers $e_i$ encode the complex structure $\tau$ of the 
torus in the usual way \cite{fk80}. 

If we use $\CP^2_{(1,1,2)}$ scaling to set $\lambda = {1 \over w}$ (or
$\lambda = {1 \over x}$ if $w=0$), we can relate the single-valued
variables $x$ and $\eta$ to the multivalued variable $s$ via the
Weierstrass $\cal P$-function with an appropriate choice of the $e_i$'s,
but the explicit form of the relation will not be needed here.  It is
nevertheless useful to know how to translate qualitatively between the
$s$-plane and the $(w,x,\eta)$ space.  First, the unique $\Z_2$
automorphism fixing a point $(s=0)$ of the torus is $s\to -s$, while the
obvious $\Z_2$ automorphism of (\ref{torus2}) is $\eta\to -\eta$ with
$w$ and $x$ fixed.  The $\Z_2$ identification $(v,s) \simeq (-v,-s)$
therefore reads $(v,w,x,\eta) \simeq (-v,w,x,-\eta)$ in $(v,w,x,\eta)$
language.  The fixed points of this map are the half-periods
$s=0,\tau/2,1/2,(1+\tau)/2$, or the branch points $x=e_i w$.  Thus, we
can choose to match these fixed points up in some way, for
example, 
\bea
(w,x,\eta)=(1,e_1,0) &\leftrightarrow & s=0,\nonumber\\ 
(w,x,\eta)=(1,e_2,0) &\leftrightarrow & s={1\over2},\nonumber\\ 
(w,x,\eta)=(1,e_3,0) &\leftrightarrow & s={\tau\over2},\nonumber\\ 
(w,x,\eta)=(1,e_4,0) &\leftrightarrow & s={1\over2}(\tau+1),
\eea
although any other ordering would do as well.
 
Next, we orbifold the space $\til Q$ under the identification 
$(v,w,x,\eta) \simeq (-v,w,x,-\eta)$ by using the single-valued variables: 
\be\label{cov} 
y \equiv v\eta, \qquad\qquad z \equiv v^2,  
\ee
(and $w$ and $x$ unchanged) so that the orbifolded background space $Q_0
= \til Q/\Z_2$ is given by the surface
\be\label{bkgrnd0} 
y^2 = z \prod_{i=1}^4 (x-e_i w) , 
\ee
in $\C\times \CP^2_{(1,1,2)}$, which follows simply by multiplying
(\ref{torus2}) by $v^2$ and changing variables as in (\ref{cov}).  At
any finite value of $z$ this curve describes a torus, but at $z=0$ this
becomes $y=0$ which is a sphere in the weighted projective space. This
can be understood by noting that for $v \neq 0$, the orbifold
$(v,w,x,\eta) \to (-v,w,x,-\eta)$ identifies two different tori, while
at $v=0$ it identifies the two $\eta$-sheets of one torus, which
effectively removes any non-trivial cycles and reduces the torus to a
sphere.  Thus the M theory lift of this IIA brane configuration can be
pictured as being similar to the snocone geometry pictured in figure~2,
with, however, the constant $v\neq0$ ($z\neq0$) cross sections being
similar tori, instead of circles, and the $v=0$ ($z=0$) cross section
being a sphere instead of a line segment.  This corresponds to the naive
construction of the M theory lift of the IIA configuration by adding
circles.
 
As a further check on this background surface, note that it is singular 
(has $\Z_2$ orbifold points) at the four points $(w,x,y,z) = 
(1,e_i,0,0)$.  It should reduce to the $D_2$ singular space in the 
vicinity of a pair of these singularities.  To see that this is indeed 
the case, take, say, $e_1=1$, $e_2=-1$ and $e_3$ and $e_4$ very large 
(which we can do for any pair of nearby points by an appropriate 
coordinate change).  Then, in the vicinity of the $e_1$ and $e_2$ 
orbifold points we can set $w=1$ (this is just choosing an appropriate 
coordinate patch in the $\CP^2$), rescale $z$ to absorb the $e_3 e_4$ 
factor, and get approximately 
\be  
y^2 = z (x^2-1) . 
\ee  
This is precisely the form of the $D_2$ singularity found in the 
literature, see \eg\ \cite{sw9607}. 

\subsection{$Q_0$: Deforming the orbifold} 
 
Now we deform this background space by (in the IIA picture) moving the 
D6 branes off the orbifold points.  This corresponds to complex 
deformations of the background surface (\ref{bkgrnd0}) which do not 
change its asymptotic (large $z$) behavior.  Physically it corresponds 
to turning on masses for the four fundamental flavors. 
 
The most general holomorphic terms that can be added to the right hand 
side of (\ref{bkgrnd0}) consistent with the surface having 
weight four in $\CP^2_{(1,1,2)}$  are of the form 
\be\label{dfrmtn} 
z^{n_2} y^2 + z^{n_1} y P_2(x,w) + z^{n_0} P_4(x,w) 
\ee 
where $P_m(x,w)$ is an arbitrary homogeneous polynomial with constant 
coefficients of weight $m$ in $x$ and $w$, and the $n_i$ are integers 
whose possible values we will now determine.  The basic requirement on 
the $n_i$ is that they do not lead to terms which are either singular 
at finite $z$ or change the topology of the surface at large $z$.  The 
first requirement implies that all the $n_i$ must be non-negative. 
The second is easily seen to imply that $n_0=0$ or $1$ while $n_1= 
n_2=0$.  Not all of these terms are deformations of (\ref{bkgrnd0}), 
though, since some can be reabsorbed in holomorphic redefinitions of the 
coordinates.  Such redefinitions which are regular as $|z|\to\infty$ 
and respect the $\CP^2_{(1,1,2)}$ structure of the background are of 
the form 
\bea\label{cptransf}
y &\to& Ay + P_2(x,w),\nonumber\\ 
x &\to& P_1(x,w),\nonumber\\ 
w &\to& P_1(x,w),\nonumber\\ 
z &\to& Bz+C , 
\eea 
where $A$, $B$ and $C$ are constants, and the $P_i(x,w)$ are again 
homogeneous constant coefficient polynomials (and the two $P_1$'s 
need not be the same).  These changes of variables can be used 
to absorb the $n_2=0$ term in (\ref{dfrmtn}) by a rescaling of $y$, 
the $n_1=0$ term by a $P_2$ shift in $y$ (completing the square), 
the $n_0=1$ term by adjusting the $e_i$ and rescaling $z$, and 
the coefficient of the $x^4$ term in $P_4$ in the $n_0=0$ term 
by a shift in $z$.  Thus we are left with precisely four complex 
deformations: 
\be\label{masses} 
y^2 = z \prod_{i=1}^4 (x-e_i w) - \tilde\mu_1 x^3 w  
- \tilde\mu_2 x^2 w^2 - \tilde\mu_3 x w^3 - \tilde\mu_4 w^4, 
\ee 
where the $\tilde\mu_i$ are the four deformation parameters (which we 
will relate to the masses below).  

Note that the only part of the (\ref{cptransf}) coordinate
redefinitions left which preserve the form of the surface
(\ref{masses}) are the three complex parameter $\SL(2,\C)$
transformations $x \to Ax+Bw$, $w \to Cx+Dw$, with $AD-BC=1$
(accompanied by appropriate rescalings or shifts of $y$ and $z$), and
the scale transformation $z \to \Delta^2 z$, $y \to \Delta y$, leaving
$x$ and $w$ fixed.  The $\SL(2,\C)$ transformations leave appropriate
combinations of the $\tilde\mu_i$ deformation parameters invariant,
but transform the $e_i$ by fractional linear transformations:
\be
e_i \to (Ae_i + B)/(Ce_i+D)
\ee
(see section~5.3 below for the details).  Thus of the four $e_i$
parameters, only one combination is coordinate invariant.  Under the
scale transformation the $e_i$ are invariant, but the deformation
parameters scale as $\tilde\mu_i \to \Delta^2 \tilde\mu_i$.  But the
normalization of an additional structure on the surface, the
Seiberg-Witten differential $\lambda$ to be constructed in the next
subsection, fixes the the scale transformation.  Thus the complex
structure of the $Q_0$ surface (\ref{masses}) is described by five
complex parameters, the four deformation parameters $\tilde\mu_i$
(which we will see have the interpretation as fundamental
hypermultiplet masses) and the invariant combination of the $e_i$
(which corresponds to a dimensionless coupling constant in the gauge
theory).

It is easy to check that turning on the $\tilde\mu_i$ indeed resolves
the four singularities at $x=e_iw$.  Rewriting (\ref{masses})
as
\be\label{masses2} 
y^2 = (z+1) \prod_{i=1}^4 (x-e_i w) -  
\prod_{i=1}^4 (x-[\delta_i +e_i]w) 
\ee 
where the $\delta_i$ are functions of the $\tilde\mu_i$ deformation
parameters, it is apparent that turning on $\delta_i$ resolves the
singularity at $x=e_i w$.
 
Another check is to compare our surface to the resolution of the $D_2$
singular space appearing in the literature.  Recall that taking $e_1=1$,
$e_2=-1$, $e_3$ and $e_4$ large, and rescaling by $e_3e_4$ reduced the
undeformed surface (\ref{bkgrnd0}) to the $D_2$ singularity.  Doing the
same to the deformed surface (\ref{masses2}) gives (in the $w=1$
coordinate patch)
\be
y^2 = z(x^2-1) + (\delta_1+\delta_2)x  
+ (\delta_1-\delta_2 -\delta_1\delta_2). 
\ee 
This is the same as the deformation of the $D_2$ singularity given in 
\cite{hov9705},\footnote{This is equation (40) of \cite{hov9705} with 
the change of variables $x \to y$, $y \to x$, and $z \to -z$.} 
$y^2 = z( x^2 -1) - 2\mu_1 \mu_2 x - (\mu_1^2 + \mu_2^2)$ 
with an appropriate map between their $\mu_i$'s and our $\delta_i$'s. 
 
\subsection{Hyperkahler structure and one-form} 
 
It follows from \cite{w9703} that the Seiberg-Witten differential 
$\lambda$ depends solely on the background M theory hyperkahler 
geometry $Q_0$ in which the M5 branes are embedded.  Here we will use 
this fact to construct $\lambda$ for the unshifted ($M=0$) background. 
In the next subsection when we turn on the shift, we will show how it 
is modified. 
 
$\lambda$ can be computed \cite{fs9706} as the solution of 
\be\label{ome1} 
\omega = d\lambda 
\ee 
where $\omega$ is a suitably chosen holomorphic (2,0)-form on the 
background space $Q_0$ given by (\ref{masses}).  $\omega$ is determined by 
the hyperkahler structure of $Q_0$ which implies that $Q_0$ has three 
independent complex structures $I$, $J$ and $K$, satisfying 
$I^2=J^2=K^2=-1$ and $IJ=K$.  If we choose the coordinates $(w,x,y,z)$ 
to be holomorphic with respect to the $I$ complex structure, then 
$\omega$ is defined by 
\be\label{hks} 
\omega = \omega_J + i \omega_K, 
\ee 
where $\omega_{J,K}$ are the Kahler forms associated with the $J$, $K$  
complex structures of $Q_0$, respectively. 
 
At constant $z$, $Q_0$ is a torus, and therefore admits the holomorphic 
one-form $(wdx-xdw)/y$. Hence the general holomorphic (2,0)-form on $Q_0$  
can be written as 
\be\label{ome2} 
\omega = f(z) {(wdx-xdw)\wedge dz\over y} 
\ee 
with $f(z)$ a polynomial in $z$ to ensure holomorphicity at $z=0$. 
$f(z)$ can be determined by consistency with (\ref{hks}) and the
metric on $Q_0$.  Specifically, as $|z|\to\infty$ we know from the 
M theory construction that the metric on $Q_0$ approximates the $\Z_2$ 
orbifold (\ref{or3}) of the flat $\C\times T^2$ space coordinatized by 
$(v,s)$.  In the $(x,\eta)$ coordinates on $T^2$ (going to a $w=1$ patch 
of $\CP^2_{(1,1,2)}$) the flat line element is 
\be 
ds^2 = dv\, d\bar v + |\alpha|^2 {dx\, d\bar x \over |\eta|^2}, 
\ee 
where $\alpha$ is a complicated constant which depends on the 
conventional way the flat $T^2$ coordinate $s$ is expressed in terms 
of $x$.  Since $\alpha$ will only enter in the overall normalization 
of $\lambda$ we need not specify its precise value. 
 
Going to the $(x,y,z)$ orbifold coordinates using (\ref{cov}) gives 
the metric at large $z$ as 
\be 
ds^2 = {dz\, d\bar z \over 4|z|}  
+ |\alpha|^2 {|z| dx\,d\bar x \over |y|^2} . 
\ee 
Using (\ref{hks}) to write $\omega_J = {1\over2}(\omega + 
\bar\omega)$, and the relation $J = g^{-1} \omega_J$ between the 
complex structure, metric, and Kahler form, we compute the nonzero 
components of $J$ at large $z$ to be 
\bea 
J^{\bar x}_z = {\bar y f\over2|\alpha|^2|z|}, &\qquad& 
J^z_{\bar x} = -{2|z|\bar f\over \bar y},  \nonumber\\ 
J^x_{\bar z} = {y \bar f\over2|\alpha|^2|z|}, &\qquad& 
J^{\bar z}_x = -{2|z| f\over y}. 
\eea 
Consistency with the hyperkahler structure of $Q_0$ implies that $J^2=-1$ 
which implies $|f|^2=|\alpha|^2$ at large $z$.  Since $f$ is polynomial 
in $z$ this fixes 
\be 
f(z)=\alpha. 
\ee 
It is convenient to choose $\alpha=1/2$ so that the final form for the 
(2,0)-form on $Q_0$ is 
\be 
\omega = {(wdx-xdw)\wedge dz\over 2y} . 
\ee 
 
We now solve (\ref{ome1}) for $\lambda$.  Since $d\omega=0$ a local 
solution always exists.  It is not too hard to see that the global 
solution is given by 
\be\label{lam2} 
\lambda = {y(xdw-wdx)\over \prod_{i=1}^4 (x-e_iw)} 
\ee
which is unique up to a total derivative. 
 
The residues of $\lambda$ at its eight poles at $x=e_iw$ yield linear 
combinations of the hypermultiplet bare masses appearing in the 
classical Lagrangian. The residues $\mu_i$ are readily obtained by 
performing contour integrals of (\ref{lam2}) about the poles $e_i$ 
yielding: 
\be 
\mu_i = {1\over2\pi i} \oint\lambda = \pm {\sqrt{y(x=e_i,w=1)}\over 
\prod_{j\neq i} (e_i-e_j) } . 
\ee 
This allows us to rewrite the deformation parameters $\tilde\mu_j$ 
of (\ref{masses}) in terms of the residues $\mu_j$ so that the 
background surface $Q_0$ becomes 
\be\label{crv1} 
y^2 = z\prod_i (x-e_iw) + \sum_j \mu_j^2 w  
\prod_{k\neq j} [(x-e_kw)(e_j-e_k)] . 
\ee 
Note that in this way of writing the surface, $w$ and $x$ are 
dimensionless, $y$ has dimension of mass, and $z$ has dimension of 
mass-squared.  Note also that the overall multiplicative factor in 
relating $\mu_j$ to the mass parameters appearing in the Lagrangian 
has been left undetermined because we did not determine the 
normalization $\alpha$ of $\lambda$, but simply chose it to a 
convenient value.  This factor will be determined in section 5 by 
matching to weak coupling limits of the theories. 

Equations (\ref{crv1}) and (\ref{lam2}) are our main results so far,
describing the complex structure and one-form of the background surface
$Q_0$. 

\subsection{$Q_M$: Implementing the shift}

Now we turn on the shift in $v$ (in the IIA picture), which, as
discussed in section 2, translates into turning on a (relative)
antisymmetric mass in the field theories.  We follow the discussion of
the shift in the elliptic models of \cite{w9703}, rephrasing the
essentials of incorporating the shift in $(s,v)$ coordinates.  We then
formulate specific conditions on the background and curve in terms of
the orbifold-invariant $(w,x,y,z)$ coordinates.  This gives the shifted
background in the undeformed ($\mu_i=0$) case.  As a final step, we show
how to deform this procedure to implement the shift in the $\mu_i\neq0$
background, which corresponds to turning on the four fundamental masses. 

Denote by $X_M$ the affine bundle over $T^2$ formed from $\til Q = T^2
\times \C$ by making the two identifications (\ref{or4}) and
(\ref{or5}), and denote by $X_M\backslash P$ the manifold $X_M$ minus
the fiber (copy of the $v$-plane) over a conveniently chosen point $P:
\{s=s_0\} \in T^2$ in the base torus.  Then, as discussed in
\cite{w9703}, the complex structure of $X_M\backslash P$ is the same as
that of $X_0\backslash P$, {\em i.e.}\ of the unshifted bundle over
$T^2$.  So we can specify the complex structure of $X_M$ by specifying
an affine analytic change of variables on the fibers
\be\label{cov2.5}
\til v = v + g(s) , 
\ee
between coordinates $(s,v)$ on $X_0\backslash P$ and the coordinates
$(s,\til v)$ on $X_M\backslash P$.  In particular, a holomorphic
function on $X_M$ is an arbitrary holomorphic function on $X_0\backslash
P$ ({\em i.e.}\ a meromorphic function on $X_0$ with singularities only
at the $P$ fiber) which is regular in the $(s,\til v)$ variables at
$s=s_0$.  As discussed in \cite{w9703}, $g(s)$ has a simple pole at $P$
with residue proportional to $M$.  Since the regularity conditions on
functions on $X_M$ need only be determined in a vicinity of $P$, we need
only specify $g$ in a neighborhood of $P$.  Thus any function with a
simple pole at $P$ will do.  It is easy to translate this prescription
to the $\CP^2_{(1,1,2)}$ coordinates $(w,x,\eta)$ on $T^2$.  The change
of variables (\ref{cov2.5}) becomes
\be\label{cov3} 
\til v = v + g(w,x,\eta) 
\ee 
where $g$ has a simple pole at $P$ with residue proportional to $M$.
Without loss of generality, we can choose to work in a $\CP^2_{(1,1,2)}$
coordinate patch with $w=1$, which we will do from now on.  If the point
$P$ has coordinates $(w,x,\eta)=(1,x_0,\eta_0)$, then one convenient
choice of the function $g$ is
\be
g = M {\eta+\eta_0 \over x-x_0} .
\ee
It is easy to check that it has a simple pole at $x=x_0$ on the sheet of
the torus where $\eta=\eta_0$ and is regular on the other sheet where
$\eta=-\eta_0$. 

We now want to orbifold this construction to obtain a description of the
shifted M theory background, which we will denote $Q_M$.  Recall that
the orbifolding makes the $\Z_2$ identification $(v,w,x,\eta) \simeq
(-v,w,x,-\eta)$, which fixes the four branch points of the torus.  We
can orbifold the above prescription directly if we take the point $P$ to
be one of the branch points.  However, it turns out to be more
convenient to choose a general point on the torus to remove.  Call this
point $P_+$ with coordinates which we take to be $(w,x,\eta) =
(1,x_0,\eta_0)$.  Now the image of this point under the $\Z_2$
identification on the torus is another point $P_-$ with coordinates
$(w,x,\eta)=(1,x_0,-\eta_0)$.  Thus to orbifold $X_M$ we need to
generalize the construction of $X_M$ of \cite{w9703} to the case where
not one but two fibers of $X_0$ are excised.   

But this is easy to do: we define the complex structure of $X_M$ by
identifying its holomorphic functions as those on $X_0\backslash \{P_+
\cup P_-\}$ which are regular in a neighborhood of the $P_\pm$ fibers
when reexpressed in terms of the new fiber coordinates
\be\label{cov3.5}
\til v_\pm = v + g_\pm(x,\eta)
\ee
where $g_\pm$ have simple poles at $P_\pm$ respectively, and the sum of
their residues is proportional to $M$.  An obvious way to respect the
$\Z_2$ orbifold symmetry is to choose
\be\label{cov3.7}
g_\pm = M {\eta\pm\eta_0\over x-x_0} .
\ee

We perform the orbifolding by rewriting these changes of variables in
terms of $\Z_2$-invariant coordinates.  Away from $x=x_0$ ($P_\pm$)
recall that these are $y=v\eta$ and $z=v^2$ (as well as $w$ and $x$
which are already $\Z_2$-invariant).  It follows from (\ref{cov3.5}) that
the $\Z_2$ action identifies the new coordinates in the neighborhoods of
the two fibers by
\be
\til v_+ \leftrightarrow -\til v_-, \qquad
\eta \leftrightarrow -\eta ,
\ee
(while $x$ and $w$ are left invariant).  Thus a basis of algebraically
independent $\Z_2$-invariant coordinates can be taken to be
\be
\zt = {1\over4} (\til v_+ + \til v_-)^2,
\qquad
\yt = {1\over2} (\til v_+ + \til v_-) \eta ,
\ee
since the other possible invariant, $\til v_+ - \til v_-$ is
just a function of $x$ by (\ref{cov3.7}), and so is not independent.
Plugging in from (\ref{cov3.5}) and (\ref{cov3.7}) then gives
the desired singular change of variables in a neighborhood of
the $x=x_0$ ``fiber'' of $Q_0$:
\bea\label{cov4}
\yt &=& y \left( 1 + M {y\over z(x-x_0)} \right) ,\nonumber\\
\zt &=& z \left( 1 + M {y\over z(x-x_0)} \right)^2 .
\eea

Note that what before the $\Z_2$ orbifold identification was two fibers
over $(x,\eta)=(x_0,\pm\eta_0)$, has become a single ``fiber'' in $Q_0$
at $x=x_0$.  This is apparent from the equation of the $Q_0$ surface
(\ref{bkgrnd0}), which we recall is
\be\label{ucrv}
y^2 =zP(x)\qquad\mbox{where}\qquad P(x)=\prod_{i=1}^4 (x-e_i),
\ee
since the two roots of $y$ at $x=x_0$ are joined at $z=0$.

The specific change of variables (\ref{cov4}) is not unique; many other
particular forms are possible since it is only the local behavior near
the $x=x_0$ ``fiber'' that is important.  This local behavior can be
made more apparent by expanding out (\ref{cov4}) using the equation of
the surface $Q_0$ (\ref{ucrv}) giving
\bea\label{cov2}
\yt &=& y + M {P(x)\over(x-x_0)} ,\nonumber\\
\zt &=& z + 2M {y\over(x-x_0)} + M^2 {P(x)\over(x-x_0)^2} .
\eea
Thus $y$ is shifted by a term with a simple pole at $x_0$ plus regular
terms, while $z$ is shifted by a double pole there plus single pole and
regular terms. 

Nevertheless, the specific change of variables (\ref{cov4}) has some
handy properties.  The inverse is given by simply changing $M$ to $-M$:
\be\label{cov4inv} 
y = \yt \left(1 - M {\yt\over\zt(x-x_0)}\right), \qquad 
z = \zt \left(1 - M {\yt\over\zt(x-x_0)}\right)^2. 
\ee 
Also, in their patch the $\yt$ and $\zt$ coordinates satisfy the same
relation defining the $Q_0$ surface as $y$ and $z$ do (\ref{ucrv}),
namely
\be
\yt^2 =\zt P(x) .
\ee
These properties will prove useful in section~4 for describing an 
explicit recipe for writing down M5 brane curves in $Q_M$, so we will strive
to preserve them in generalizing the above change of variables to the
case where the D6 brane (fundamental mass) deformations $\mu_i$ are turned on.

The generalization of the shift to the case with the $\mu_i$
deformations turned on is straightforward.  Recall that the equation
(\ref{crv1}) for the mass-deformed $Q_0$ surface is
\be\label{crv1.5}
y^2 = z P(x) + Q(x)
\ee
where
\be
Q(x) = \sum_j \mu_j^2 \prod_{k\neq j} [(x-e_k)(e_j-e_k)] ,
\ee
$P(x)$ is as in (\ref{ucrv}), and $j,k=1,\ldots,4$.  For general
$\mu_i$ the deformation only significantly changes the complex
structure of $Q_0$ in a ($z$-dependent) neighborhood of the $x=e_i$
branch points or for small $z$ (compared to the $\mu_i^2$).  Therefore
the only region of $Q_0$ where the change of variables (\ref{cov4})
near $x=x_0$ might break down is for small $z$.  But, since in a small
enough neighborhood of the $x=x_0$ ``fiber'' the surface $Q_0$ is
smooth whether or not the $\mu_i$ vanish, in fact the change of
variables (\ref{cov4}) applies to the deformed case without change.

However, it will be useful to let the change of variables depend on the
deformation parameters in such a way as to preserve the handy features
mentioned above.  When the $\mu_i=0$, the two branches of the $x=x_0$
``fiber'' join at $z=0$ as noted above; for non-zero $\mu_i$ it follows
from (\ref{crv1.5}) that they now join at $z= - Q(x_0)/P(x_0)$.  This
suggests that a natural way to generalize (\ref{cov4}) is to
shift $z$ to $z +Q(x)/P(x)$, giving
\bea\label{cov5}
\yt &=& y \left( 1 + M {y\over \left(z+{Q(x)\over P(x)}\right)
(x-x_0)} \right) ,\nonumber\\
\zt +{Q(x)\over P(x)}&=& \left(z + {Q(x)\over P(x)} \right)
\left( 1 + M {y\over \left(z+{Q(x)\over P(x)}\right) (x-x_0)} \right)^2 .
\eea
It is easy to check that this form of the change of variables has the
properties that its inverse is given just by interchanging
$y\leftrightarrow\yt$ and $z\leftrightarrow\zt$ and changing $M\to-M$;
and that the new variables $(x,\yt,\zt)$ in the $x\sim x_0$ patch
satisfy the same equation (\ref{crv1.5}) for the unshifted $Q_0$
surface.  Furthermore, since near $x=x_0$ (\ref{cov5}) differs from
(\ref{cov4}) just by a constant shift in $z$, it has the same
analyticity properties.  In fact, expanding (\ref{cov5}) out
using (\ref{crv1.5}) gives back precisely (\ref{cov2}).

Finally, a short computation shows that the one-form $\lambda$ has the
same form (\ref{lam2}) in both the $y$, $z$ and the $\yt$, $\zt$
coordinates, and so is not modified in the presence of the shift $M$.
Though $M$ is interpreted as a mass, it does not appear as the residue
of a pole in $\lambda$ on $Q_M$.  Instead, the curve describing how
the NS5 brane lies in $Q_M$ will have poles (\ie, will go off to
infinity in $Q_M$) inducing poles in the restriction of $\lambda$ to
this curve.  The mass $M$ will thus appear as the residue of the
induced pole at $x=x_0$.  The same is also true of other masses in the
theory, such as the bifundamental hypermultiplet masses, which appear
as residues of induced poles in $\lambda$ corresponding to the asymptotic
infinities of the NS5 branes in the IIA picture.

In summary, we have found that we can implement the shift in the
background surface (whether deformed or not) from $Q_0$ given by
(\ref{crv1.5}) to $Q_M$ through the change of variables (\ref{cov2}) in
the neighborhood of an arbitrary point $x_0$.  In particular, in
section~4.4 we will write curves for M5 branes on $Q_M$ by writing them
as curves on $Q_0$ with singularities at $x=x_0$ such that they are
regular at $x=x_0$ when written in terms of the $(w,x,\yt,\zt)$
variables defined in (\ref{cov2}). 

\section{M5 brane curve}
 
By lifting the D6 branes and O6$^-$ planes of the IIA string
configuration described in section~2 to M theory, we have (in
section~3) derived the complex structure of the M theory background
geometry.  This geometry is a deformed $\Z_2$ orbifold of the product
of the $v$-plane and the $x$-$y$ torus ($\C \times T^2$), with a shift
in $v$ along one of the torus cycles.  We now want to embed an M5
brane curve corresponding to the D4 and NS5 branes in this background.
First, in section~4.1, we explain the form of the simple curves
describing the M theory lift of isolated D4 branes, NS5 branes, NS5
branes stuck at an orbifold point, and intersecting D4 and NS5 branes.
With this formulaic dictionary under our belts, we then proceed to
writing down the general curve describing the intersection of many 4
and 5 branes.  We do this following the three steps introduced in
section~3: in section~4.2 we write the curve in the undeformed
background and impose orbifold invariance on it; in section~4.3 we
determine how the curve must be modified in the mass-deformed
background; in section~4.4 we implement the shift.

\subsection{M5 brane lifts of NS5 and D4 branes}

\noindent{\em NS5 brane lifts}
\smallskip

In the geometry of section~2 an NS5 brane lies at constant $x_6$ and
is extended along the $v$ directions.  Its M theory lift (in the
non-orbifolded case---the elliptic model of \cite{w9703}) thus is just
the fiber above a point $(w,x,\eta)=(1,x_0,\eta_0)$ on the torus:
\be
x=x_0\qquad\mbox{and}\qquad \eta=\eta_0 .
\ee
(Here and henceforth we are working in a $w=1$ coordinate patch in
$\CP^2_{1,1,2}$; thus $\eta_0$ is one of the roots of $\eta_0^2 =
P(x_0)$ with $P(x_0)$ given in (\ref{ucrv}).)

Now consider the corresponding curve in the orbifolded model of the
undeformed $Q_0$ background (\ie\ two O6$^-$ planes each with a pair
of coincident D6 branes).  The orbifold-invariant variables are $x$,
$y=v\eta$, and $z=v^2$, so the curve for an NS5 brane at $x=x_0$
becomes simply
\be\label{ns5crv}
x=x_0.
\ee
Note that since on the orbifolded background
\be\label{u2crv}
y^2 = z P(x) \qquad \mbox{where}\quad P(x)=\prod_{i=1}^4 (x-e_i) ,
\ee
there is a qualitative difference between an NS5 brane at a generic
point $x=x_0$, and a ``stuck'' 5 brane at one of the four branch
points $x=e_i$.  This is apparent from the $z\neq 0$ behavior of these
curves: the generic (unstuck) NS5 brane is at two values of $y$ for a
given $z$, while the stuck brane is at only one (namely $y=0$).  Thus
an unstuck NS5 brane can be thought of as forming a double cover of
the $z$-plane, while a stuck NS5 brane ony covers it once.  This
corresponds to the fact that the unstuck brane carries twice the
charge of a stuck brane.  Thus, despite appearances, making an
arbitrarily small change in the 5 brane curve (\ref{ns5crv}) from
$x=e_i$ to $x=e_i + \delta x$ is in fact a singular transformation.

In this way of describing the M theory lift of NS5 branes the
existence of stuck NS5 branes looks dependent on the existence of
orbifold points.  This is not actually the case, as we can see by
deforming the orbifold by moving the D6 branes off the O6$^-$ planes
(or, equivalently, by turning on the $\mu_i$ masses).  Recall the
surface (\ref{crv1}) for the mass-deformed $Q_0$ background:
\be\label{dcrv}
y^2 = z P(x) + Q(x) 
\ee
where
\be
Q(x) = \sum_j \mu_j^2 \prod_{k\neq j} [(x-e_k)(e_j-e_k)] .
\ee
Clearly the curve $x=x_0$ for a generic (unstuck) NS5 brane lift
in the undeformed case retains the same topology and asymptotics
when set in the deformed background:  it still forms a double
cover of the $z$-plane, and at large $z$ its $y$ coordinate
values asymptote to the same values as in the undeformed case.

The lift of a stuck brane, $x=e_i$, however, has a qualitatively
different behavior in the deformed background.  It describes two {\em
disconnected} single covers of the $z$ plane, each at constant $y =
\pm\sqrt{Q(e_i)} = \pm \mu_i \prod_{k\neq i} (e_i-e_k)$.  In order to
keep the asymptotic topology unchanged (\ie\ that it be a single
cover of the $z$ plane) we must choose one or the other of these two
solutions as the deformation of the stuck NS5 brane.  (We do not really
have to think of these as two distinct deformations of the stuck NS5
brane since once we have chosen one, the other can be realized by
changing the phase $\mu_i \to e^{i\pi} \mu_i$ of the relevant
deformation parameter.) In summary, a stuck NS5 brane in the
mass-deformed $Q_0$ background is described by a pair of equations
\be
x=e_i \quad\mbox{and}\quad y=y_i \equiv \sqrt{Q(e_i)} ,
\ee
and covers the $z$ plane once.  It is stuck since it admits no
continuous deformation, since for any $x\neq e_i$ there are only the
topologically distinct unstuck NS5 branes which are double covers of the
$z$ plane.

\smallskip
\noindent{\em D4 brane lifts}
\smallskip

A D4 brane lies at constant $v$ and is extended along $x_6$.  Its M
theory lift in the (non-orbifolded) elliptic model thus wraps the
$x_6$-$x_{10}$ torus, and so is described in the $(v,x,\eta)$
coordinates of the $\C\times T^2$ transverse space simply by
\be
v = v_0 .
\ee
Orbifolding under $v\to -v$ by placing an image brane at $v=-v_0$,
implies that the M5 brane lift of a D4 brane in the orbifold-invariant
$(x,y,z)$ coordinates of (\ref{u2crv}) becomes simply
\be\label{d4crv}
z=z_0.
\ee
This has no asymptotic region, but wraps the $x$-$y$ torus once.  

There is also a ``stuck'' version of the lift of a D4 brane, namely
the one at $z=0$, which has the topology of a sphere---the $y=0$
subspace of $\CP^2_{(1,1,2)}$---which passes through all four orbifold
points of the $Q_0$ background.  Upon mass-deforming the $Q_0$
background, this curve is deformed to $z=-Q(x)/P(x)$ in order to
retain its spherical topology on the $x$-$y$ torus, since it still has
$y=0$ as its solution.\footnote{More general deformations are
possible: $z=(R^2-Q)/P$ where $R$ is any degree two polynomial in $x$
which vanishes as the deformation is turned off also has the spherical
topology.  The above argument still applies, however.}  But this curve
has poles (infinities) in $z$ at the $e_i$, and so cannot be
considered a small deformation of the stuck D4 brane no matter how
small the $\mu_i$.  So the stuck D4 brane only exists in the $\C\times
T^2/\Z_2$ orbifold geometry, and should be considered as a part of a
special degeneration of a generic D4 brane.  This agrees with the
conclusion reached in section~2 based on the IIA brane geometry, and is
also implicitly supported by \cite{ll9708}, where stuck D4
branes were not found in the M theory lift of a single O6$^-$, the
Atiyah-Hitchin space.

\smallskip
\noindent{\em D4-NS5 intersections}
\smallskip

Now we would like to describe the intersection of a D4 brane with an
unstuck NS5 brane in a neighborhood of the 5 brane.  In the IIA limit
they are described by $(z-z_0)(x-x_0)=0$.  This is the equation for two
curves intersecting transversely at two points since for $x=x_0$ and
$z=z_0$ there are two values of $y$ satisfying (\ref{u2crv}).  But in
the M theory lift these intersections will typically be smoothed out.
So the question is what are the complex deformations of these
intersections which preserve the topology in the vicinity of (but not
at) the intersection points, and also have the same asymptotics along the
NS5 brane far from the intersection points?  Since an unstuck NS5 brane
$x=x_0$ is described by two values of $y$ for each $z\neq0$, preserving
the topology means that for each $x\neq x_0$ (but close enough to $x_0$)
there should be precisely two solutions for $(y,z)$ also satisfying
(\ref{u2crv}).  Having the right asymptotics means that for large $z$
the curve should approach the unstuck NS5 brane curve $x=x_0$.  There
are only two possible complex deformations of the intersection
satisfying these conditions:
\be\label{defint}
(z-z_0)(x-x_0)= \epsilon + \delta y
\ee
for arbitrary complex $\epsilon$ and $\delta$.  The $\epsilon$ term
smooths out intersection and clearly does not affect the asymptotics.
That the $\delta$ term does not either is less obvious; but since for
fixed $x$, $y\propto\sqrt z$ by (\ref{u2crv}), the $\delta$ term is
subleading for large $z$.  Finally, it is not too hard to show that for
any $\epsilon$ and $\delta$, an $x$ close enough to $x_0$ can be found for
which there are always two solutions to (\ref{defint}) and
(\ref{u2crv}). 

The intersection of a D4 ($z=z_0$) with a stuck NS5 ($x=e_i$ and
$y=y_i\equiv \sqrt{Q(e_i)}$), however, has only a single deformation:
\be\label{stuckint}
(z-z_0)(x-e_i) = \delta (y+y_i),
\ee
for arbitrary complex $\delta$.  This is because it is only for this
form that the asymptotics of the stuck NS5 brane are preserved.  In
particular, for fixed large $z$ there is only one solution to
(\ref{stuckint}) and (\ref{dcrv}), and so (\ref{stuckint}) describes
asymptotically in a neighborhood of $x=e_i$ only a single-sheeted cover
of the $z$ plane, which is moreover asymptotically close to $y=y_i$.
This is easiest to see by rewriting (\ref{stuckint}) as
\be\label{sknt}
z = \delta {y+y_i\over x-e_i} + z_0 .
\ee
Then large $z$ occurs only at poles of the right side which are at
$x=e_i$ and $x=\infty$.  (Recall that $x=\infty$ is a regular point,
corresponding to finite $x$ with $w=0$ in a different coordinate patch
of $\CP^2_{(1,1,2)}$.) But the pole at $x=\infty$ is not in a
neighborhood of the asymptotic NS5 brane, so should be discarded in this
local argument.  There are generically two poles at $x=e_i$ since there
are two points on the (fixed $z$) torus with this coordinate:
$(x,y)=(e_i,\pm y_i)$.  However, precisely because of the form of the
numerator in (\ref{sknt}) the pole at $(e_i,-y_i)$ is cancelled,
leaving only a simple pole at $(e_i,+y_i)$, as desired.

The form of the curve for the smoothed intersection of a NS5 brane and
a D4 brane stuck at $z=0$ also bears discussion.  In the $\C\times
T^2/\Z_2$ background, following the prescription developed above, the
curve is just (\ref{defint}) with $z_0=0$, which, as before, should be
thought of as being valid in the patch of the $x$-$y$ torus away from
$x=\infty$.  For the special case of $\epsilon=0$, even though this
curve does not factorize, it actually describes two intersecting
curves.  This follows since not only is there the usual solution in
which $z$ blows up as $x$ approaches $x_i$, but there is also the
solution $z=0$ for all $x$ since by (\ref{u2crv}) $y=0$ if $z=0$
independent of $x$.  These two branches of solutions intersect at the
orbifold points $(x,y,z) =(e_j,0,0)$, and it is not hard to check that
near the intersections the two branches are described by $z\propto
x-e_j$ and $z=0$, which, in terms of the single-valued local
complex coordinate $\xi = \sqrt{x-e_j}$, describes a tangent
intersection of two surfaces.  In constrast, the curve describing a
generic D4-NS5 intersection, \ie\ (\ref{defint}) with either of
$\epsilon$ or $z_0$ non-zero, or with the mass deformation $Q(x)$ in
(\ref{dcrv}) turned on, describes only a single curve.  This is
another way of seeing that the stuck D4 brane only exists in the $\C
\times T^2/\Z_2$ orbifold geometry.

\subsection{M5 brane curve in $Q_0$} 

The previous subsection described the M5 brane lifts of intersecting NS5
and D4 branes in the $Q_0$ background geometry.  However this
description was only local in the compact $x$-$y$ (``torus'')
directions.  The global problem of writing a curve with a given number
$q$ of stuck NS5 branes at asymptotic (large $z$) positions $x=e_i$ and
$r$ of unstuck NS5 branes at asymptotic postions $x=x_i$ in $Q_M$ is the
subject of the rest of this section.  We start in this subsection by
considering such curves in the $\C\times T^2/\Z_2$ orbifolded background and
then include the mass deformations of the $Q_0$ space, and in the next
subsection we include the shift of the general $Q_M$ space. 

A convenient starting point for the construction of the curve for $q$
stuck and $r$ unstuck NS5 branes and $k$ D4 branes in the orbifolded
background is to consider the curve for Witten's elliptic model
\cite{w9703} with $q+2r$ NS5 branes and $k$ D4 branes.  To respect the
orbifold invariance\footnote{We are working for the rest of this section
in the $w=1$ $\CP^2_{(1,1,2)}$ coordinate patch.} $(v,x,\eta) \to
(-v,x,-\eta)$ we must distribute $q$ of the 5 branes among the various
branch points $(x,\eta)=(e_p,0)$, and put the rest at the $r$ pairs of
points $(x_s,\pm\eta_s)$.  We will denote the coordinates of all these
points by $(x_I,\eta_I)$ for $I=1,\ldots,q+2r$, and will assume without
loss of generality that none of the $x_I=\infty$. 

Having located the NS5 branes in M theory, we can write down the
elliptic model curve in terms of the unorbifolded coordinates
$(v,x,\eta)$ following \cite{w9703}:
\be\label{curve1} 
0 = F(v,x,\eta) = v^k + \sum_{\ell=1}^k v^{k-\ell} f_\ell(x,\eta) .
\ee
The $f_\ell$ can have at most simple poles at the locations of the NS5
branes on the torus.  The most general such functions can be written as
\be\label{coeffs} 
f_\ell(x,\eta) = \til A_\ell + \sum_{I=1}^{q+2r} \til B_{I\ell} 
{\eta + \eta_I \over x-x_I}
\ee 
where $\til A_\ell$ and $\til B_{I\ell}$ are complex constants 
constrained by
\be
\sum_I \til B_{I\ell} = 0 \qquad\mbox{for all}\quad\ell.
\ee
This last constraint is needed to prevent a pole at $x=\infty$.

Next, we arrange the curve symmetrically with respect to the O6$^-$
plane reflection $(v,x,\eta) \to (-v,x,-\eta)$:
\be\label{cur1}
F(v,x,\eta) = F(-v,x,-\eta). 
\ee
Since the coefficient of the $v^k$ term is one, this can only be
satisfied for $k$ even.  This condition implies
\be\label{co2}
f_\ell(x,-\eta) = (-)^\ell f_\ell(x,\eta), 
\ee
and it follows from (\ref{coeffs}) that 
\be\label{co3} 
f_\ell = \left\{ \matrix{ 
A_\ell + \sum_{s=1}^r {B_{s\ell} \over x-x_s}
& \qquad\ell\ \mbox{even}\cr
\eta \left( \sum_{s=1}^r {C_{s\ell} \over x-x_s}
+ \sum_{p=1}^q {D_{p\ell} \over x-e_p} \right)
& \qquad\ell\ \mbox{odd}\cr 
} \right. 
\ee 
where $A_\ell$, $B_{s\ell}$, $C_{s\ell}$, and $D_{p\ell}$ are arbitrary
complex coefficients (formed from combinations of the (\ref{coeffs})
coefficients $\til A_\ell$, $\til B_{I\ell}$ and $\eta_I$) and subject
to the constraint that
\be
\sum_{s=1}^r C_{s\ell} + \sum_{p=1}^q D_{p\ell} =0 ,\qquad
\mbox{for all odd}\ \ell.
\ee

We can now write the curve in terms of the orbifold-invariant variables
$x$, $y=v\eta$ and $z=v^2$.  The results are
\be\label{keven} 
0=z^n + A(z) + \sum_{s=1}^r { B_s(z) + y C_s(z)
\over x-x_s } + \sum_{p=1}^q { y D_p(z) \over x-e_p }
\qquad \mbox{for}\quad k=2n ,
\ee
where $A$, $B_s$, $C_s$, and $D_p$ are arbitrary polynomials in $z$ of
order $n-1$, subject to the constraints that
\be
\sum_{s=1}^r C_s(z) + \sum_{p=1}^q D_p(z) =0
\qquad \mbox{for all}\quad z .
\ee
 
It might be thought that $k$ odd solutions in which $F(-v,x,-\eta)
=-F(v,x,\eta)$ should also be allowed.  For example, a single D4
brane stuck at the orbifold invariant point $v=0$ clearly describes
an orbifold invariant submanifold even though its equation changes sign.
Allowing this behavior, and then expressing $vF(v,x,\eta)$ in terms of
the orbifold-invariant variables gives
\be
0 = z^n + z \til A(z) + \sum_{s=1}^r { z\til B_s(z) + y C_s(z)
\over x-x_s } + \sum_{p=1}^q { y D_p(z) \over x-e_p }
\qquad \mbox{for}\quad k=2n-1 , 
\ee
where now $\til A(z)$ and $\til B(z)$ are arbitrary polynomials of
order $n-2$ in $z$.  This is just a specialization of the $k$-even
curve, realized by setting the constant terms of $A(z)$ and $B_s(z)$
to zero.  This is precisely the specialization of the generic D4-NS5
intersection discussed at the end of the last subsection: the constant
and $z^0 (x-x_s)^{-1}$ terms are set to zero.  But as discussed there,
this specialization does not survive any deformation of the
background, and so should not be counted as a separate case.  So we
will consider only $k$ even from now on.

In the case with two stuck NS5 branes (the $q=2$ cases of section 2),
the brane picture allows an odd number of D4 branes (corresponding to
having $\SU$ gauge factors of odd rank).  Since the naive $k$ odd
solutions were ruled out above, the question arises as to how to
construct them when $q=2$.  As we will show in section 5.4, in
precisely the $q=2$ case of two stuck 5 branes, a decoupling limit
which embeds the scale-invariant $\SU(2n-1)$ theory in the scale-invariant
$\SU(2n)$ theory is possible, and gives the $k$ odd curve.  It
differs from the $k$ even curve essentially by just deleting the 
leading $z^n$ term in (\ref{keven}).

Next, we determine how the curve must be deformed once we deform the
orbifold space to $Q_0$ by turning on masses as in (\ref{dcrv}).  First,
note that we cannot change the order of the curve in $z$, since this
corresponds to changing the number of D4 branes in the type IIA picture.
Similarly, we must keep the same number of simple poles in the
coefficients of each power of $z$ corresponding to the locations of the
NS5 branes.  Finally, from the discussion in the last subsection, it
follows that the positions of the $q$ stuck NS5 branes must be kept the
same, \ie\ at $x=e_p$, but that their numerators should be shifted from
$y$ to $y-y_p$ where
\be\label{ypdef}
y_p \equiv \sqrt{Q(e_p)}.
\ee
It follows that the curve in the $Q_0$ background is
\be\label{cur2} 
0=F(x,y,z) \equiv
z^n + A(z) + \sum_{s=1}^r { B_s(z) + y C_s(z)
\over x-x_s } + \sum_{p=1}^q { (y-y_p) D_p(z) \over x-e_p }
\qquad \mbox{for}\quad k=2n .
\ee
 
It might seem that the choice of sign of the square root in
(\ref{ypdef}) for each $p$ give us many distinct curves.  However it
is not too hard to see that the different choices of sign correspond
to the same theory albeit at different values of its parameters.
For we can always write the deformed background in the form
\be
y^2 = (z-1) \prod_{i=1}^4 (x-e_i) - \prod_{i=1}^4 (x-e_i-\delta_i), 
\ee
where the $\delta_i$ are functions of the $\mu_i$.  When the
$\delta_i$ vanish the branch points of the torus (at a given $z$) are
at the points $x_i = b_i$ with $b_i = e_i$.  When the $\delta_i$ are
all small but non-zero, a power series expansion for the branch points
$b_i$ gives
\be\label{biei}
b_i = e_i + {1\over z} \delta_i + {\cal O}(\delta^2). 
\ee
Now in the curve (\ref{cur2}) the choice was whether to put the pole at
a given $x=e_i$ on one sheet or the other.  But by (\ref{biei}) as
$\delta_i \to e^{2\pi i}\delta_i$, the position of the pole traverses a
complete circle around the branch point in the $x$-plane, thus shifting
from one sheet to the other.  Thus we learn that we can go from one
choice of any one sign of the square root in (\ref{ypdef}) to the other
by a continuous change in the deformation parameters.  This was anticipated 
in the IIA brane picture by a deformation argument at the end of
section~2.1. 

\subsection{M5 brane curve in $Q_M$} 
 
Now we generalize the results of the previous subsection from the
non-shifted ($M=0$) case to the case of an arbitrary shift parameter
$M$. As discussed in section~3.4, the curve for the shifted model is
like the curve for the non-shifted model except that in addition to
the poles at $x=x_s$ and $x=e_p$ in (\ref{cur2}), we should also allow
singularites at $x=x_0$ (the excised ``fiber" of $Q_0$), subject to the
constraint that the curve be regular in a coordinate patch covering
$x=x_0$ with coordinates $(x,\yt,\zt)$ related to $(x,y,z)$ as in
(\ref{cov2}). 

Precisely which order singularities should be allowed in (\ref{cur2})
can be determined by the form of the change of variables (\ref{cov2})
to the $(x,\yt,\zt)$ coordinate patch which we repeat here:
\bea\label{bicov}
\yt &=& y + M {P(x)\over(x-x_0)} ,\nonumber\\
\zt &=& z + 2M {y\over(x-x_0)} + M^2 {P(x)\over(x-x_0)^2} .
\eea
Recalling that the inverse change of variables is of the same form with
$y\leftrightarrow\yt$, $z\leftrightarrow\zt$ and $M\to-M$, it follows
that the inverse change of variables can only introduce or remove poles
at $x=x_0$ or $x=\infty$, and since $F(x,y,z)$ in (\ref{cur2}) should be
regular at $x=x_0$ when written in the $(x,\yt,\zt)$ coordinates, the
general form of the transformed $F$, $\til F(x,\yt,\zt)$, must be that
of (\ref{cur2}) but with possible extra poles at infinity:
\be\label{Ftil}
0=\til F(x,\yt,\zt) \equiv
\zt^n + \sum_{\alpha=0}^\infty 
\left[\til A_\alpha(\zt) + \yt \til E_\alpha(\zt)\right] x^\alpha
+ \sum_{s=1}^r {\til B_s(\zt) + \yt \til C_s(\zt) \over x-x_s} + 
\sum_{p=1}^q {(\yt-y_p) \til D_p(\zt) \over x-e_p}
\ee
where the $\til A_\alpha$, $\til B_s$, $\til C_s$, $\til D_p$, and $\til
E_\alpha$ are arbitrary polynomials of order $n-1$ in $\zt$.  (The
additional terms compared to (\ref{cur2}) are the $\til A_{\alpha\neq0}$
and $\til E_\alpha$ polynomials which parametrize the possible
singularities at $x=\infty$.)  Now substituting in the change of
variables (\ref{bicov}) into (\ref{Ftil}) and using the $Q_0$ equation
$y^2 = z P(x) + Q(x)$ to reexpress all higher powers of $y$ in terms of
polynomials in $x$, $y$, and $z$ that are at most linear in $y$, the
result is a curve of the form (\ref{cur2}) but with additional poles at
$x_0$.  Computing the order of these poles is then a simple matter of
counting.  Since the change of variables (\ref{bicov}) for $\zt$ is of
the form $\zt=z+P_1 y + P_2$ and $\yt=y+P_1$, where $P_i$ refer to
rational functions of $x$ with poles of up to order $i$ at $x_0$, then
just counting powers of $y$, $z$, and the order of poles at $x_0$ gives
\bea
\zt^\ell &\sim& z^\ell+\sum_{a=1}^\ell z^{\ell-a}
(P_{2a} + y P_{2a-1}) ,\nonumber\\
\yt \zt^\ell &\sim& \sum_{a=0}^\ell z^{\ell-a}
(P_{2a+1} + y P_{2a}) .
\eea
Since the highest power of $\zt$ alone is $\ell=n$, and of $\yt\zt^\ell$ is
$\ell=n-1$ in (\ref{Ftil}), it follows that the highest order pole at
$x=x_0$ in the coefficient of a term with $z^{n-\ell}$ is $2\ell$, while
for that of a term with $yz^{n-\ell}$ it is $2\ell-1$. 

Thus, the general form of the curve with $k$ D4 branes, $q$ stuck NS5
branes, and $r$ unstuck NS5 branes in $Q_M$ is
\be\label{QMcur} 
0= z^n + A(z,x) + y E(z,x) + \sum_{s=1}^r { B_s(z) + y C_s(z) \over x-x_s } 
+ \sum_{p=1}^q { (y-y_p) D_p(z) \over x-e_p }
\qquad \mbox{for}\quad k=2n ,
\ee
where 
\be
y_p \equiv \sqrt{Q(e_p)} ,
\ee
$A$, $E$, $B_s$, $C_s$, and $D_p$ have expansions
\bea
A(z,x) &=& \sum_{\ell=1}^n z^{n-\ell} \sum_{a=0}^{2\ell} A_{a\ell}
(x-x_0)^{-a} ,\nonumber\\
E(z,x) &=& \sum_{\ell=1}^n z^{n-\ell} \sum_{a=1}^{2\ell-1} E_{a\ell}
(x-x_0)^{-a} ,\nonumber\\
B_s(z) &=& \sum_{\ell=1}^n z^{n-\ell} B_{s\ell} ,\nonumber\\
C_s(z) &=& \sum_{\ell=1}^n z^{n-\ell} C_{s\ell} ,\nonumber\\
D_p(z) &=& \sum_{\ell=1}^n z^{n-\ell} D_{p\ell} ,
\eea
and satisfy the constraint
\be\label{constr9}
E_{1\ell} + \sum_{s=1}^r C_{s\ell} + \sum_{p=1}^q D_{p\ell} = 0
\qquad\mbox{for all}\ \ell
\ee
(so that there is no pole at $x=\infty$).  Demanding regularity at
$x=x_0$ after making the change of coordinates
\bea\label{vocib}
y &=& \yt - M {P(x)\over(x-x_0)} ,\nonumber\\
z &=& \zt - 2M {\yt\over(x-x_0)} + M^2 {P(x)\over(x-x_0)^2}, 
\eea
which is just the inverse of (\ref{bicov}), then determines all the
$A_{a\ell}$ and $E_{a\ell}$ in terms of $A_{0\ell}$, $B_{s\ell}$, 
$C_{s\ell}$, and $D_{p\ell}$.  This curve
(\ref{QMcur}) lives on the $Q_M$ surface whose equation (excluding the
$x=x_0$ submanifold) we repeat here:
\bea\label{final}
y^2 &=& z P(x) + Q(x), \nonumber\\
P &\equiv& \prod_{i=1}^4 (x-e_i), \nonumber\\
Q &\equiv& \sum_{i=1}^4 \mu_i^2 \prod_{j\neq i} [(x-e_j)(e_i-e_j)] . 
\eea
Finally, the $x$ and $y$ directions are compactified by two points at
infinity, which can be realized by embedding all curves in
$\CP^2_{(1,1,2)}$---\ie\ by adding powers of a new variable $w$ to
each term to make all terms homogenous in $(w,x,y)$ with weights
assigned as $(1,1,2)$, respectively.  This completes our construction
of the general M5 brane curve in the $Q_M$ background; this paragraph
assembles all the relevant formulas.

The following sections will discuss explicit examples of these curves.

\section{Consistency checks}
 
In this section we perform some simple consistency checks on our
curves.  The number of such possible checks is immense: there should
be an intricate web of relations among these curves and other curves
known in the literature corresponding to the relations among the $N=2$
gauge theories as gauge factors are decoupled (by taking weak coupling
limits), or Higgsed at weak coupling (by taking vevs large), or as
matter hypermultiplets are decoupled (by taking masses large).  We
will only describe here a sample of a few of the simplest checks,
though it will be apparent that many more can quickly and easily be
performed.  The main point of the examples will be to illustrate how
to handle the weak coupling and large mass limits, and how to deduce
the appropriate identifications of the parameters appearing in the
curves with the classical masses, couplings, and vevs.  These
identifications are good semi-classically, but are necessarily
ambiguous up to non-perturbative redefinitions \cite{ap9911}.  They
are deduced by taking various weak coupling limits of the theory in
which various patterns of gauge group breakings and matter decouplings
are reproduced.  

Along the way we will also show that the background $Q_M$ surface, when
thought of as a curve at fixed $z$ is equivalent to the curve describing
the Coulomb branch of the scale invariant $\SU(2)$ theory with four
fundamental hypermultiplets.  Also, in section 5.4 we will show how 
the curves for $\SU(k)$ theories with $k$ odd can be derived by a simple 
breaking procedure.

\subsection{Parameter counting} 

The most basic check is that the number of parameters entering into
the curve (\ref{QMcur})---(\ref{final}) match the number of parameters
describing the corresponding gauge theory on its Coulomb branch.  The
background surface $Q_M$ has six parameters: $M$, the four $\mu_i$,
and the one $\SL(2,\C)$-invariant combination of the $e_i$.  The curve
(\ref{QMcur}) has $r$ NS5 brane positions $x_s$ as well as the
independent coefficients of the $A$, $B_s$, $C_s$, and $D_q$
functions.  The regularity conditions at $x=x_0$ under the coordinate
change (\ref{vocib}) imply that the coefficients of inverse powers of
$(x-x_0)$ can be expressed in terms of the leading coefficients (\ie,
those with no $x-x_0$ dependence).  Since each of $A$, $B_s$, $C_s$
and $D_p$ are polynomials of order $n-1$ in $z$, they each have $n$
independent leading coefficients, for a total of $n(1+2r+q)$.  However
(\ref{constr9}) enforces $n$ relations among them.  Thus the total
number of parameters is $6+r+2rn+qn$.  Recalling that $k=2n$ is the
number of D4 branes, this matches precisely the counting of parameters
given at the end of section 2 for the gauge theories.  (The counting
also works for the degenerate $q=r=0$ case, as the reader can easily
check.)

\subsection{Some examples}

Let us now concentrate on two simple cases to illustrate parameter
matching.  We will focus on the ($r=0$, $q=1$, $k=2n$) and the ($r=0$,
$q=2$, $k=2n$) curves.

\bigskip
\noindent{($r=0$, $q=1$, $k=2n$) curve}
\smallskip

According to the IIA brane correspondence described in section 2, this
case describes the Coulomb branch of the $\Sp(2n)$ gauge theory with 4
fundamental and one antisymmetric hypermultiplet.  Putting the one
stuck NS5 brane at $x=e_1$, and placing (for simplicity) the shift at
$x_0=0$, the curve reads
\be
0 = z^n + A(z,x) + y E(z,x) + {y-y_1 \over x-e_1} D(z) 
\ee
where 
\bea
A &=& \sum_{\ell=1}^n z^{n-\ell} \sum_{a=0}^{2\ell} A_{a\ell} x^{-a}, 
\nonumber\\
E &=& \sum_{\ell=1}^n z^{n-\ell} \left( -D_\ell x^{-1} + 
\sum_{a=2}^{2\ell-1} E_{a\ell} x^{-a} \right), \nonumber\\
D &=& \sum_{\ell=1}^n z^{n-\ell} D_{\ell} , \nonumber\\
y_1 &=& \sqrt{Q(e_1)},
\eea
and the $A_{a\ell}$, $E_{a\ell}$ and $D_{\ell}$ coefficients must be 
chosen so that all poles at $x=0$ cancel upon changing to $(\yt,\zt)$ 
variables according to
\be\label{fivethree}
y = \yt - M {P(x)\over x} ,\qquad
z = \zt - 2M {\yt\over x} + M^2 {P(x)\over x^2}, 
\ee
where $P(x)$, $Q(x)$ are given in (\ref{final}).

For example, when $n=2$, we have
\bea
A &=& z\left( A_{01} + {A_{11}\over x} + {A_{21}\over x^2} \right)
+ A_{02} + {A_{12}\over x} + {A_{22}\over x^2} + {A_{32}\over x^3} 
+ {A_{42}\over x^4} ,\nonumber\\
E &=& - z {D_1\over x} - {D_2\over x} + {E_{22}\over x^2} + {E_{32}\over x^3},
\nonumber\\
D &=& z D_1 + D_2 ,
\eea
and we find, after a second of CPU time,
\bea
A_{11} &=& 6 M^2 (s_3 - e_1 s_2)
\nonumber\\
A_{21} &=& 6 M^2 e_1 s_3
\nonumber\\
A_{12} &=& A_{01} M^2 (s_3 - e_1 s_2) + 4 q_1 M^2 + 8 q_0 M^2 e_1^{-1} 
+ 4 y_1 M^3 s_2
\nonumber\\
&& \qquad {}
+ 4 M^4 \left[ e_1^2 (s_1 s_2 - s_3) - e_1 s_2^2 + s_3 s_2  \right]
\nonumber\\
A_{22} &=& A_{01} M^2 e_1 s_3 + 4 q_0 M^2 - 4 y_1 M^3 s_3
+ M^4 \left[ e_1^2 (s_2^2 - 4 s_3 s_1) 
+ 2 e_1 s_3 s_2 - 3 s_3^2 \right]
\nonumber\\
A_{32} &=& 2 M^4 e_1 s_3 (s_3-e_1 s_2)
\nonumber\\
A_{42} &=& M^4 e_1^2 s_3^2 
\nonumber\\
D_1 &=& -4 M
\nonumber\\
D_2 &=& -2 A_{01} M + 8 M^2 y_1 e_1^{-1} + 4 M^3 (2 e_1 s_1 - s_2)
\nonumber\\
E_{22} &=& 4 M^3 (2 s_3 - e_1 s_2)
\nonumber\\
E_{32} &=& 4 M^3 e_1 s_3
\eea
where we have defined
\bea
q_0 = Q(0), \quad &&
q_1 = Q'(0), \nonumber\\
s_1 = e_2 + e_3 + e_4, \quad &&
s_2 = e_2 e_3 + e_2 e_4 + e_3 e_4, \qquad
s_3 = e_2 e_3 e_4 ,
\eea
and $A_{01}$ and $A_{02}$ are arbitrary.

Note that when we send the shift parameter to zero, $M\to 0$, only the
$A_{0\ell}$ coefficients remain non-zero, as in the curves we
found in section 4.2.  Thus, in this limit the
curve reduces to
\be\label{masslessSP}
0 = z^n + A(z) = z^n + \sum_{\ell=1}^n z^{n-\ell} A_{0\ell}
\ee
which is precisely the form of the $r=q=0$ curve.  This bears out the
interpretation of the $r=q=0$ curve given in section 2 as the
$\Sp(2n)$ theory with four fundamentals and a massless antisymmetric.

Furthermore, the curve for this massless antisymmetric theory was
found in \cite{asty9611,dls9612} where it was shown that the Coulomb
branch is generically the tensor product of $n$ copies---one for each
independent $\Sp(2n)$ adjoint scalar vev---of the one-dimensional
Coulomb branch of the $\SU(2)$ with four fundamental flavors theory.
But that is precisely the content of (\ref{masslessSP}): its solutions
are just the $n$ roots $z_s$, $s=1,\ldots,n$, of the polynomial
$z^n+A(z)$.  When plugged back into the equation (\ref{final}) for the
$Q_0$ surface this gives the $n$ tori
\be\label{fiveeight}
y^2 = z_s P(x) + Q(x)
\ee
where $P(x)$ and $Q(x)$ are given as usual by (\ref{final}).  To show
the equivalence with the known $\Sp(2n)$ curve, it therefore just
remains to show the equivalence of the curve (\ref{fiveeight}) (which
is just the $Q_0$ surface (\ref{final}) at fixed $z$) with the curve
\cite{sw9408} for the $\SU(2)$ theory with four fundamentals.  This
requires matching the coupling, masses, and vev parameters, and will
be done in section 5.3 below.

\bigskip
\noindent{($r=0$, $q=2$, $k=2n$) curve}
\smallskip

According to the IIA brane correspondence described in section 2, this
case describes the Coulomb branch of the $\SU(2n)$ gauge theory with
4 fundamental and 2 antisymmetric hypermultiplets.  Putting the
two stuck NS5 branes at $x=e_2$ and $x=e_3$, and placing (for simplicity) 
the shift at $x_0=0$, the curve reads
\be
0 = z^n + A(z,x) + y E(z,x) + {y-y_2 \over x-e_2} D_1(z) + 
{y-y_3 \over x-e_3} D_2(z) 
\ee
where
\bea
A &=& \sum_{\ell=1}^n z^{n-\ell} \sum_{a=0}^{2\ell} A_{a\ell} x^{-a}, 
\nonumber\\
E &=& \sum_{\ell=1}^n z^{n-\ell} \left( -[D_{1\ell}+D_{2\ell}] x^{-1} + 
\sum_{a=2}^{2\ell-1} E_{a\ell} x^{-a} \right), \nonumber\\
D_i &=& \sum_{\ell=1}^n z^{n-\ell} D_{i \ell} , \nonumber\\
y_i &=& \sqrt{Q(e_i)},
\eea
and the $A_{a\ell}$, $E_{a\ell}$ and $D_{i\ell}$ coefficients must be 
chosen so that all poles at $x=0$ cancel upon changing to $(\yt,\zt)$ 
variables according to (\ref{fivethree}).

For example, when $n=2$, we have
\bea
A &=& z\left( A_{01} + {A_{11}\over x} + {A_{21}\over x^2} \right)
+ A_{02} + {A_{12}\over x} + {A_{22}\over x^2} + {A_{32}\over x^3} 
+ {A_{42}\over x^4} ,\nonumber\\
E &=& - z {D_{11}+D_{21}\over x} - {D_{12}+D_{22}\over x} 
+ {E_{22}\over x^2} + {E_{32}\over x^3}, \nonumber\\
D_1 &=& z D_{11} + D_{12}, \nonumber\\
D_2 &=& z D_{21} + D_{22},
\eea
and we find, after a few more seconds of CPU time,
\bea
A_{11} &=& 3 D_{11} M e_1e_4(e_2-e_3) 
- 6 M^2 (e_1e_2e_3 - e_1e_2e_4 +e_1e_3e_4 + e_2e_3e_4)
\nonumber\\
A_{21} &=& 6 M^2 e_1e_2e_3e_4
\nonumber\\
A_{12} &=& 2 D_{11} q_0 M \left({1\over e_2}-{1\over e_3}\right) 
+ D_{22} M e_1e_4(e_2-e_3)
\nonumber\\
&& \quad {}
- A_{01} M^2 (e_1e_2e_3 + e_1e_2e_4 - e_1e_3e_4 + e_2e_3e_4) 
\nonumber\\
&& \quad {}
+ 4 q_1 M^2 + 8 q_0 M^2 e_3^{-1} 
- D_{11} M^2 y_2(e_1e_3+e_1e_4+e_3e_4)
\nonumber\\
&& \quad {}
+ D_{11} M^2 y_3 e_3^{-1} (e_1e_2e_3 + 2e_1e_2e_4 - e_1e_3e_4 + e_2e_3e_4)
\nonumber\\
&& \quad {}
+ 4 M^3 y_3 e_3^{-1} (e_1e_2e_3 + 2e_1e_2e_4 - e_1e_3e_4 + e_2e_3e_4)
\nonumber\\
&& \quad {}
- D_{11} M^3 (e_2-e_3) \bigl[ 2e_2e_3(e_1^2+e_4^2) 
\nonumber\\
&& \qquad\qquad\qquad {}
+ e_1e_4(e_1e_2-2e_1e_3-e_1e_4+2e_2e_3+e_2e_4-2e_3e_4)\bigr]
\nonumber\\
&& \quad {}
+ 4 M^4 \left[ (e_1+e_4)(e_2^2e_3^2+e_1e_3e_4(e_2+e_3))
- (e_1^2+e_4^2)e_2e_3(e_2-e_3)\right]
\nonumber\\
A_{22} &=& A_{01} M^2 e_1e_2e_3e_4 + 4 q_0 M^2 + D_{11} M^2 e_1e_4(y_2e_3-y_3e_2)
- 4 y_3 M^3 e_1e_2e_4 
\nonumber\\
&& \quad {}
+ D_{11} M^3 e_1e_4(e_2-e_3)(e_1e_2e_3-e_1e_2e_4-e_1e_3e_4+e_2e_3e_4)
\nonumber\\
&& \quad {}
+ M^4 \bigl[ e_1^2e_2^2e_3^2 - e_1^2e_2^2e_4^2 + e_1^2e_3^2e_4^2 + e_2^2e_3^2e_4^2 
\nonumber\\
&& \qquad\qquad {}
+ 2 e_1e_2e_3e_4 (e_1e_2 - e_1e_3 + e_1e_4 - e_2e_3 + e_2e_4 - e_3e_4) \bigr]
\nonumber\\
A_{32} &=& D_{11} M^3 e_1^2e_2e_3e_4^2(e_2-e_3)
- 2 M^4 e_1e_2e_3e_4(e_1e_2e_3 - e_1e_2e_4 +e_1e_3e_4 + e_2e_3e_4)
\nonumber\\
A_{42} &=& M^4 e_1^2e_2^2e_3^2e_4^2  
\nonumber\\
D_{21} &=& -4 M -D_{11}
\nonumber\\
D_{22} &=& - D_{12} -2 A_{01} M 
- 2 D_{11} M \left( {y_2\over e_2} - {y_3\over e_3} \right)
+ 8 M^2 y_3 e_3^{-1} 
- 3 D_{11} M^2 (e_1+e_4)(e_2-e_3)
\nonumber\\
&& \quad {}
- 4 M^3 (e_1e_2 - 2e_1e_3 + e_1e_4 - 2e_2e_3 + e_2e_4 - 2e_3e_4)
\nonumber\\
E_{22} &=& 3 D_{11} M^2 e_1e_4(e_2-e_3)
- 4 M^3 (e_1e_2e_3 - 2e_1e_2e_4 +e_1e_3e_4 + e_2e_3e_4)
\nonumber\\
E_{32} &=& 4 M^3 e_1e_2e_3e_4
\eea
where we have defined $q_0 = Q(0)$, $q_1 = Q'(0)$, and $A_{01}$,
$A_{02}$, $D_{11}$, and $D_{12}$ are arbitrary.

Finally, when the shift parameter $M$ vanishes, the curve dramatically
simplifies to
\be\label{masslessSU}
0 = z^n + A(z) + \left( {y-y_2 \over x-e_2} - {y-y_3 \over x-e_3} \right)
D(z)
\ee
where now $A$ and $D$ are arbitrary polynomials of degree $n-1$ in $z$.

\subsection{Identifying coupling and mass parameters: $\Sp$ theory} 

The gauge coupling and theta angle of the scale invariant theory can
be combined into a complex coupling which we denote
\be
q = e^{2\pi i\tau} ,
\ee
where $\tau = (\theta/2\pi) + i (4\pi/g^2)$.  Likewise, denote the
fundamental hypermultiplet masses\footnote{We use an unconventional
normalization of the hypermultiplet masses $m$ so that they enter the
classical action as ${\cal W} = \sqrt2 \{\mbox{tr}(Q\Phi\til Q) + m
\mbox{tr}(Q\til Q) \}$ where ${\cal W}$ is the superpotential for the
the $N=2$ theory written in $N=1$ notation with $(Q,\til Q)$ the two
$N=1$ chiral superfields making up the $N=2$ hypermultiplet, and
$\Phi$ the $N=1$ adjoint chiral superfield part of the $N=2$ vector
multiplet.  Canonically normalized masses $\til m$ are then given by
$\til m = \sqrt2 m$.} that appear in the weak coupling action by
$m_i$.  In this subsection we will find the relation between the gauge
coupling $q$ and masses $m_i$ and the parameters $e_i$ and $\mu_i$ of
the $Q_M$ surface.  We will do this first for the $\Sp(2n)$ theories
(or, more generally, for theories with only one stuck NS5 brane), and
leave the $\SU(k)$ case for the next subsection.

First, however, it will be useful in what follows to record the
explicit change of variables on $y$ and $z$ needed to keep the form of
the surface the same under $\SL(2,\C)$ transformations on $w$ and $x$.
(Recall the discussion after equation (\ref{masses}) in section 3.2).
Write the surface in terms of the same variables with tildes, and
redefine them by
\bea\label{sl2c}
\xt &=& Ax+Bw\nonumber\\ 
\wt &=& Cx+Dw\nonumber\\ 
\yt &=& \left[\prod_i (A - \et_i C)\right] y 
\eea 
with $AD-BC=1$.  The inverse of this change of variables is 
\bea\label{tran1} 
x &=& D\xt-B\wt\nonumber\\ 
w &=& -C\xt+A\wt\nonumber\\ 
y &=& \left[\prod_i (C e_i +D)\right] \yt . 
\eea 
Here $e_i$ and $\et_i$ are related by 
\be 
\et_i = {Ae_i+B\over Ce_i+D},\qquad 
e_i = {D\et_i-B\over-C\et_i+A}, 
\ee 
which imply the useful identities 
\be 
A-\et_i C = (C e_i+D)^{-1}, 
\ee 
and 
\be 
\et_i-\et_j = (e_i-e_j)(Ce_i+D)^{-1}(Ce_j+D)^{-1} . 
\ee 
Plugging this change of variables into the tilded surface gives back
the original surface after some algebra and with $z$ related to $\zt$
by
\be\label{tran2} 
z = \left[\prod_i(Ce_i+D)\right] \zt + C 
\sum_j {\mu_j^2 \over Ce_j +D} \prod_{k\neq j} (e_j-e_k) . 
\ee 

We can now identify the gauge coupling parameter $q$ in terms of the
parameters of the $Q_M$ background.  General arguments \cite{ap9911}
imply that $q$ can be a function only of dimensionless parameters, so
we can set the mass parameters $\mu_i=0$.  Then the background
geometry is $y^2=z\prod_i(x-e_i)$, which at fixed $z$ describes a
torus with ($\SL(2,\C)$-invariant) complex structure $\tau_t$.  In the
IIA brane picture the imaginary part of $\tau_t$ measures the
circumference of the $x_6$-circle.  Since the inverse-squared gauge
coupling is the distance between the O6$^-$ planes in the brane
picture, we identify $\tau = {1\over2}\tau_t$, or
\be\label{qgqtrel} 
q = 4\sqrt{q_t} 
\ee 
where $q_t \equiv e^{2\pi i\tau_t}$.  The factor of 4 in
(\ref{qgqtrel}) is for later convenience; it represents a modification
to the definition of the coupling by some one-loop threshold
correction \cite{ap9911}.  In general, the above identification of $q$
could be modified by adding terms with higher powers of $\sqrt q_t$
to the right hand side: these would just represent a non-perturbative
redefinition of the coupling which does not affect the weak coupling
limit $q\to0$.  In this section matching parameters in the weak coupling 
limit is all we will be interested in, since that is sufficient to take
decoupling limits.  More detailed checks against instanton expansions
require a more refined matching of parameters, as in \cite{ap9911}.

According to \cite{sw9408}, $q_t$ is related to the $e_i$, in the
$q_t\to0$ limit, by
\bea 
e_1 &=& -(1/3) + 8\sqrt{q_t} + {\cal O}(q_t) \nonumber\\ 
e_2 &=& -(1/3) - 8\sqrt{q_t} + {\cal O}(q_t) \nonumber\\ 
e_3 &=& +(2/3) + {\cal O}(q_t) \nonumber\\ 
e_4 &=& \infty . 
\eea 
It will be convenient in what follows to make an $\SL(2,\C)$
transformation, $e_i \to (Ae_i +B)/(Ce_i +D)$, which takes the average
of $e_1$ and $e_2$ to $0$, $e_3\to1$, and $e_4\to-1$.  The
transformation which does this is (to order $q_t$)
\be 
\pmatrix{A&B\cr C&D\cr} = {1\over 3\sqrt2} \pmatrix{-3&-1\cr3&-5\cr}, 
\ee 
and gives the new $e_i$ as 
\bea\label{newes} 
e_1 &=& - 4\sqrt{q_t} = -q , \nonumber\\ 
e_2 &=& + 4\sqrt{q_t} = +q , \nonumber\\ 
e_3 &=& +1 , \nonumber\\ 
e_4 &=& -1 . 
\eea 
Then the weak coupling limit corresponds to the limit in which
the torus degenerates by having the $e_1$ and $e_2$ branch points
collide.  In terms of the IIA brane picture, we thus see that
$e_1$ and $e_2$ correspond to the lift of one O6$^-$ plane, while
$e_3$ and $e_4$ are the lift of the other one.

We now turn to identifying the $\mu_i$ deformation parameters with
the fundamental hypermultiplet masses $m_i$.  The masses and deformation
parameters, by dimensional analysis and the fact that the residues of 
the Seiberg-Witten one-form can only depend linearly on the masses 
with no $q$-dependence, must therefore be related by 
\be\label{mmu} 
\mu_i=a_{ij}m_j  
\ee  
for some matrix of constants $a_{ij}$.\footnote{Since the
Seiberg-Witten one-form found in section 3.3 above could be multiplied
by an arbitrary function of $q$, the $a_{ij}$ constants may actually be
functions of $q$.  We only determine them in the $q\to0$ limit here.}
We can determine this matrix by the following symmetry argument.  On
the world-volume of 2 D6 branes in the presence of an O6${}^-$ plane
we expect an enhanced $\SO(4)$ gauge symmetry when they all coincide.
Therefore the global flavor symmetry which should be manifest in our
background---which is found by deforming two pairs of D6 branes away
from two O6${}^-$ planes---should be $\SO(4)\times \SO(4)$.  (The full
theory will have an $\SO(8)$ flavor symmetry; but this need not be
manifest in our particular way of writing the background surface.)
Therefore the $\mu_i^2$ parameters that appear in the background
surface should be the $\SO(4)\times \SO(4)$ adjoint invariants,
thinking of the masses as the eigenvalues of a mass matrix $m^i_j$ in
the adjoint.  For one $\SO(4)$ factor such invariants are tr$(m^2)$
and $\mbox{Pf}(m)$ (the Pfaffian of $m^i_j$), which in terms of its
eigenvalues $m_1$ and $m_2$ are $m_1^2 + m_2^2$ and $m_1 m_2$
respectively. A similar story applies to the second $\SO(4)$ as well
(with mass eigenvalues $m_3$ and $m_4$).  Since $\mu_1$ and $\mu_2$
are associated with the resolution of one O6${}^-$ plane
(corresponding to the branch points at $e_1$ and $e_2$, say) they are
therefore functions of the $m_1$ and $m_2$ masses only; likewise
$\mu_3$ and $\mu_4$ are functions of $m_3$ and $m_4$ only.  Thus we
expect
\be
\mu_1^2 = a(m_1^2+m_2^2) + b(m_1m_2),\qquad
\mu_2^2 = c(m_1^2+m_2^2) + d(m_1m_2), 
\ee
for some numbers $a$, $b$, $c$, and $d$, and similarly for $m_3$ and
$m_4$.  For this to be consistent with (\ref{mmu}), and using the
symmetry between $\mu_1$ and $\mu_2$, and the fact that the BPS masses 
$m_i$ should be integer linear combinations of the residues $\mu_i$ 
(corresponding to the number of times their cycles encircle the poles)
implies
that
\be\label{phmasses}
\mu_1 = \textstyle{1\over2}(m_1 + m_2) , \qquad 
\mu_2 = \textstyle{1\over2}(m_1 - m_2) , \qquad
\mu_3 = \textstyle{1\over2}(m_3 + m_4) , \qquad
\mu_4 = \textstyle{1\over2}(m_3 - m_4) . 
\ee
 
So, with (\ref{newes}) and (\ref{phmasses}) we have 
an explicit parameterization of the $Q_M$ surface (\ref{final}): 
\bea\label{crv3} 
y^2 &=& z(x^2-q^2w^2)(x^2-w^2) \\ 
&&\quad {}+ (1-q^2) \Bigl\{  
-q^2(m^2_1+m^2_2)w^2(x^2-w^2)  +2q (m_1m_2)wx(x^2-w^2) \nonumber\\ 
&&\qquad\qquad\qquad {} 
+ (m^2_3+m^2_4)w^2(x^2-q^2w^2) +2 (m_3m_4)wx(x^2-q^2w^2)\Bigr\}.
\nonumber 
\eea
The asymmetrical way in which the four fundamental masses enter in
(\ref{crv3}) is the inevitable consequence of choosing a specific weak
coupling parametrization as we have done above.  Reparametrizations of
the surface under global holomorphic coordinate changes
on $\CP^2_{(1,1,2)}$ of the form (\ref{sl2c}) together with
appropriate shifts of $z$ can be found which leave the form of
(\ref{crv3}) unchanged except for permutations of the $m_i$; see
section 5.5 below.

So far we have implicitly set the shift parameter $M$ to zero.  As
described in section 2, the shift parameter $M$ is the mass $m_A$ of
the antisymmetric hypermultiplet.  {}From the IIA brane picture $m_A$
will enter along with the $m_i$ in the identification of the $\mu_i$
parameters given in (\ref{phmasses}) above.  The reason is that the
existence of an NS5 brane stuck at one of the O6$^-$ planes breaks the
manifest $\SO(4)$ flavor symmetry at that orientifold to $\U(1)^2$.
This means that the $m_A$ mass is allowed to enter as a common shift
in the $m_1$, $m_2$ masses (if we put the stuck 5 brane at $e_1$ or
$e_2$).  Thus the $\mu_1$ and $\mu_2$ identifications of
(\ref{phmasses}) will be modified to
\be\label{antimassshift1}
\mu_1 = m_A - \textstyle{1\over2}(m_1+m_2) , \qquad
\mu_2 = \textstyle{1\over2}(m_2-m_1) .
\ee

The rest of the parameters in the curve for the 5 brane in the $Q_M$
background describe the vevs for the $\Sp(2n)\to \U(1)^n$ breaking on
the Coulomb branch.  For example, from the IIA brane picture in the
simple case where the antisymmetric mass vanishes, $M=0$, the
parameters of the resulting curve (\ref{masslessSP}) are related to
the eigenvalues $\pm\phi_\ell$, $\ell=1,\ldots,n$ of the complex
adjoint scalar field in the vector multiplet by
\be\label{Akzero}
A_{0 k} = \sum _{\ell_1 < \cdots < \ell_k}  
\phi_{\ell_1}^2 \cdots \phi_{\ell_k}^2  
\ee 
so that the roots of the curve are at
\be
z_\ell = - \phi_\ell^2.
\ee
Note that, as discussed above for the coupling and masses, these
identifications may be modified by terms with higher powers of $q$.
In addition, these identifications are also valid only for large
$\phi_\ell$: they may be modified even at ``tree level'' by shifts
proportional to powers of the masses \cite{ap9911}.  As we will
compute in the $M=0$ case momentariy, such a shift indeed occurs, and
the more accurate identification of parameters implies that the
$\phi_\ell^2$'s in (\ref{Akzero}) are replaced by $\phi_\ell^2 \to
\phi_\ell^2 - m_3^2 - m_4^2$, or equivalently, $z$ in the 5 brane
curve (\ref{masslessSP}) is shifted by
\be\label{zzshift}
z \to z - m_3^2 - m_4^2 .
\ee

Finally, we are now in a position to check the equivalence of the
$Q_0$ curve (\ref{crv3}) at fixed $z=\phi^2$ with the curve
\cite{aps9505}
\be
y^2 = (x^2-\til\phi^2)^2 - \til q \prod_{i=1}^4 (x+\til m_i) 
\ee
of the $\SU(2)$ theory with four fundamental flavors.  An explicit
$\SL(2,\C)$ transformation relating the two curves is very hard to
find, so we will make a less direct argument.  Consider the leading
$q$ behavior of the discriminants in $x$ of the right hand sides of
the two curves.  For the background curve (\ref{crv3}) it is
proportional to
\be
q^2 (z-m_3^2-m_4^2)^2 (z-m_3^2-m_4^2+m_1^2)(z-m_3^2-m_4^2+m_2^2)
(z-m_4^2)(z-m_3^2)
\ee
while for the $\SU(2)$ curve it is proportional to
\be
\til q^2 \til\phi^4 (\til\phi^2-\til m_1^2)(\til\phi^2-\til m_2^2)
(\til\phi^2-\til m_3^2)(\til\phi^2-\til m_4^2) ,
\ee
which match with the identifications (already derived above)
\be\label{spsumatch}
q \propto \til q,\qquad
z = -\til\phi^2+m_3^2+m_4^2,\qquad
m_i = \til m_i .
\ee
Since we only kept the leading terms in $q$ in the discriminants, they
miss some of the structure of the singularities of the curves at
values of $|z|\ll m_i^2$ which should also match.  We can check that
this structure also matches by comparing the structure of the curves
in the limit of large masses.  We will perform this check in section
5.5 below.

\subsection{Identifying coupling and mass parameters: $\SU$ theory}

\FIGURE{
\epsfig{file=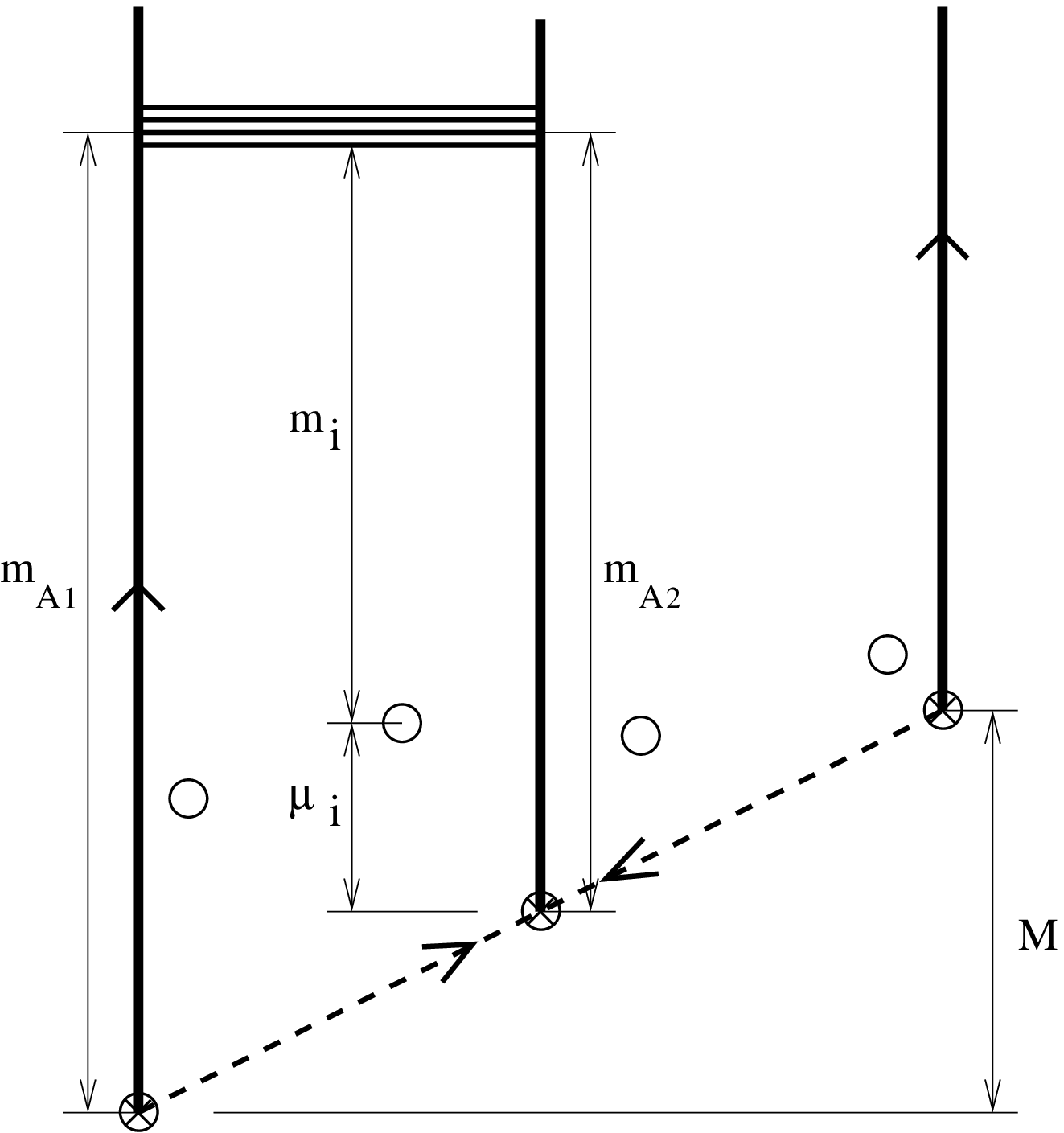,width=20em}
\caption{The brane arrangement (in the notation of figure 1) for
the $\SU(4)$ theory with two antisymmetrics and four fundamentals,
showing the brane separations corresponding to the various
hypermultiplet masses, as well as the background parameters
$\mu_i$ and $M$.}}

The matching analysis for the $\SU(k)$ theory---or for theories with
$q=2$ stuck NS5 branes, more generally---is much the same as for the
$\Sp$ ($q=1$) theories given above.  The main difference is the 
appearance of NS5 branes stuck on the two O6$^-$ planes which implies 
that the manifest $\SO(4)$ flavor symmetry at each O6$^-$ is broken to
$\U(1)^2$.  This means in practice that the relation between the
deformation parameters $\mu_i$ and the fundamental masses $m_i$
derived in the last subsection can be shifted by $\SO(4)$ singlet
masses.  {}From the IIA brane picture, summarized in figure 3, it is
easy to see that the background parameters are related to the
hypermultiplet masses by
\bea\label{antimassshift2}
\mu_1 &=& m_{A1} - \textstyle{1\over2}(m_1+m_2) , \nonumber\\
\mu_2 &=& \textstyle{1\over2}(m_2-m_1) ,\nonumber\\
\mu_3 &=& m_{A2} - \textstyle{1\over2}(m_3+m_4) , \nonumber\\
\mu_4 &=& \textstyle{1\over2}(m_4-m_3) ,\nonumber\\
M &=& m_{A1} - m_{A2} ,
\eea
where $m_{Ai}$ denote the two antisymmetric hypermultiplet masses.
Plugging these identifications along with the coupling constant
identification (\ref{newes}) into (\ref{final}) gives an explicit
parametrization of the $Q_M$ background surface for the $\SU(k)$
theories.

The rest of the parameters in the 5 brane curve describe the
eigenvalues $\phi_\ell$, $\ell=1,\ldots, 2n$, of the $\SU(2n)$ adjoint
scalar vevs on the Coulomb branch, as well as the average
antisymmetric mass
\be
m_A \equiv \textstyle{1\over2} (m_{A1} + m_{A2}) .
\ee
At least in the case where the shift parameter $M$ vanishes (so that
the two antisymmetrics have equal mass), the resulting $\SU(2n)$ curve
(\ref{masslessSU}) is easy to parameterize explicitly:
\be\label{crv2n} 
0=\sum_{a=0}^n r_{2a} (z-m_3^2-m_4^2)^{n-a} +  
{\cal P}(w,x,y) \sum_{a=1}^n r_{2a-1} (z-m_3^2-m_4^2)^{n-a} 
\ee 
where 
the $r_a$ coefficients are related (at weak coupling) to the
eigenvalues $\phi_\ell$ by
\be\label{radefn}
r_a = \sum_{\ell_1 < \cdots < \ell_a}  
(-\phi_{\ell_1}-m_A)\cdots(-\phi_{\ell_a}-m_A), 
\ee
(and $r_0\equiv 1$) where the $\phi_\ell$ satisfy $\sum_\ell \phi_\ell
= 0$, and 
\bea\label{PPshift}
{\cal P}(w,x,y) &=& {y-y_2 w^2 \over w(x-qw)} - {y-y_3 w^2 \over
w(x-w)} - (1-q^2) (m_3+m_4) \\
&=& -{(1-q)y\over(x-qw)(x-w)} -(1-q^2) \left[ {q(m_2-m_1) w \over x-qw} 
- {2m_A w - (m_3+m_4)x \over x-w } \right] .\nonumber
\eea
Note that in (\ref{crv2n}) we have shifted $z$ by (\ref{zzshift}) and
also shifted ${\cal P}$ by the constant $-(1-q^2) (m_3+m_4)$ relative
to its value given in (\ref{masslessSU}).  Both these shifts could be
undone by appropriate redefinitions of the $\phi_\ell$, which is just
to say that, as in the $\Sp$ case, these shifts reflect the
appropriate matching of the $r_a$ parameters with the weak couplings
vevs $\phi_\ell$.  Unfortunately, we know of no simpler way of
deducing these shifts than actually performing the matching to some
known Coulomb branch physics, \eg, at weak coupling.\footnote{The
$-(m_3+m_4)$ shift in (\ref{PPshift}) was determined only to leading
order in $q$; the $(1-q^2)$ factor in the shift was inserted only to
make the second line in (\ref{PPshift}) prettier.}  A simple check is
to take the curve for $n=1$ and $m_A=0$.  This should give the known
curve for the $\SU(2)$ theory with four fundamentals.  But in this
case (\ref{crv2n}) reduces to $z=\phi_1^2 + m_3^2 + m_4^2$.  Plugging
this in the background curve (\ref{crv3}) then indeed gives precisely
the $\SU(2)$ curve as we saw above in (\ref{spsumatch}).

We are now in a position to answer the question raised at the end of
section 4.2 as to what are the curves for the $q=2$ theories with an
odd number $k$ of D4 branes, corresponding to the $\SU(k)$ theories
with $k$ odd.  The $\SU(2n-1)$ theory can be found from the $\SU(2n)$
theory by going out on the Coulomb branch in a direction corresponding
to one $\phi_\ell$ getting large, while all the others remain close
together.  In terms of the eigenvalues $\phi_\ell$ and the
antisymmetric masses $m_A$ this corresponds to taking the limit
$\phi_1+m_A \equiv {\cal M} \to \infty$ keeping all the other
$\phi_i+m_A$ fixed (for $i\neq1$).  To leading order in ${\cal M}$ this
has the effect on the $r_a$ coefficients of $r_a \to {\cal M} \til
r_{a-1}$ where $\til r_a$, $a=1,\ldots,2n-1$ are the corresponding
coefficients for the $\SU(2n-1)$ theory.  Plugging these into the
curve (\ref{crv2n}) and rescaling by an overall factor of ${\cal M}$
gives, in the ${\cal M}\to \infty$ limit the new curve
\be\label{crv2n-1}
0=\sum_{a=1}^n \til r_{2a-1} (z-m_3^2-m_4^2)^{n-a} +  
{\cal P}(x,y,z) \sum_{a=1}^n \til r_{2a-2} (z-m_3^2-m_4^2)^{n-a} .
\ee
This is thus the curve for the $\SU(2n-1)$ theory; it differs
from the $\SU(2n)$ curve essentially by deleting the leading
$z^n$ term.

\subsection{Decoupling fundamental hypermultiplets} 

We now examine the decoupling of the fundamental flavors.  To decouple
a hypermultiplet we should send its mass $m_i$ to infinity while
taking the gauge coupling $q\to0$ keeping the strong coupling scale
$\Lambda=qm_i$ of the resulting asymptotically free theory fixed.

This works with no other tunings for $m_1$ and $m_2$ in (\ref{crv3})
but not so obviously for $m_3$ or $m_4$.  However, by an $\SL(2,\C)$
transformation with
\be
\pmatrix{A&B\cr C&D\cr} = \pmatrix{0&i\sqrt{q}\cr  
i/\sqrt{q} &0\cr}, 
\ee 
we can exchange $e_{1,2}$ with $e_{3,4}$, or equivalently exchange 
$m_{1,2}$ with $m_{3,4}$ in (\ref{crv3}).  Thus we can figure out the 
appropriate rescalings and shifts of $w$, $x$, $y$, and $z$ from the 
changes of variables (\ref{tran1}) and (\ref{tran2}), which are, for 
the above $\SL(2,\C)$ transformation, 
\bea\label{tran3} 
w &=& -i\xt/\sqrt{q}\nonumber\\ 
x &=& -i\sqrt{q}\wt\nonumber\\ 
y &=& \yt\nonumber\\ 
z &=& \zt + (1-q^2) 
(-m_1^2 -m_2^2 +m_3^2 +m_4^2) . 
\eea 
(This applies to the $\Sp$ or $\SU$ cases with the antisymmetric masses
set to zero;  for non-zero antisymmetric masses the $m_i$ should be
shifted according to (\ref{antimassshift1}) or (\ref{antimassshift2}).)
Since the decoupling limit of, say, $m_1$ keeps $w$, $x$, $y$, and 
$z$ fixed, so the decoupling limit of say $m_3$ should keep $\wt$, 
$\xt$, $\yt$, and $\zt$ fixed.  This implies, by (\ref{tran3}) that in 
doing the decoupling of $m_3$: 
\be\label{limit} 
m_3 =\Lambda/q,\qquad q \to 0, 
\ee 
we should scale or shift the $w$, $x$, $y$ and $z$ coordinates by 
\bea\label{shifts} 
w &\to& (1/\sqrt{q}) w\nonumber\\ 
x &\to& \sqrt{q} x\nonumber\\ 
y &\to& y\nonumber\\ 
z &\to& z + (1-q^2)(-m_1^2 -m_2^2 +m_3^2 +m_4^2). 
\eea 
Plugging the shifts (\ref{shifts}) into our surface and taking the
limits (\ref{limit}) indeed gives a well-defined surface with $m_3$
decoupled.  In this way it is easy to find changes of variables to
successively decouple all the masses.  (And once you have this, you
can forget all the above $\SL(2,\C)$-ing.)

So, explicitly: Start with our surface (\ref{crv3}) with $z$ shifted
by
\be
z \to z + (1-q^2)(m_3^2 +m_4^2) 
\ee 
so that it reads 
\bea\label{crv4m} 
y^2 &=& z(x^2-q^2w^2)(x^2-w^2) \\ 
&&\quad{}+ (1-q^2) \Bigl\{  
-q^2(m^2_1+m^2_2)w^2(x^2-w^2) 
+2q (m_1m_2)wx(x^2-w^2) \nonumber\\ 
&&\qquad\qquad\qquad{} 
+   (m^2_3+m^2_4)x^2(x^2-q^2w^2) 
+2  (m_3m_4)wx(x^2-q^2w^2)\Bigr\}.\nonumber 
\eea
Decouple $m_1$ by taking $m_1 = \Lambda_1/q$ and $q\to0$ (and no other
rescalings) giving the surface
\bea\label{crv3m} 
y^2 &=& z x^2(x^2-w^2) +
\Bigl\{-\Lambda^2_1 w^2(x^2-w^2) 
+2\Lambda_1m_2 wx(x^2-w^2)\nonumber\\ 
&&\qquad\qquad\qquad\qquad{} 
+ (m^2_3+m^2_4) x^4 
+2 m_3m_4 wx^3 \Bigr\}. 
\eea 
Next decouple $m_2$ by taking $m_2 = \Lambda_2^2/\Lambda_1$ and
$\Lambda_1\to0$ (and no other rescalings) giving the surface
\be\label{crv2m} 
y^2 = z x^2(x^2-w^2) +
\left\{2\Lambda^2_2 wx(x^2-w^2) 
+ (m^2_3+m^2_4) x^4 
+2 m_3m_4 wx^3 \right\}. 
\ee
Next decouple $m_3$ by taking $m_3 = \Lambda_3^3/\Lambda_2^2$ and
$\Lambda_2\to0$ and rescale $w = \Lambda_2^{-1} \hat w$, $x =
\Lambda_2 \hat x$, giving the surface
\be\label{crv1m} 
y^2 = -z {\hat x}^2 {\hat w}^2 +
\left(-2{\hat w}^3{\hat x} 
+ \Lambda^6_3 {\hat x}^4 
+2 \Lambda^3_3m_4 {\hat w}{\hat x}^3 \right). 
\ee
Finally, decouple $m_4$ by taking $m_4 = \Lambda_4^4/\Lambda_3^3$ and
$\Lambda_3\to0$ (and no other rescalings) giving the surface
\be\label{crv0m} 
y^2 = -z {\hat x}^2 {\hat w}^2 - 2 {\hat w}^3{\hat x}
+ 2 \Lambda^4_4 {\hat w}{\hat x}^3 . 
\ee
These five surfaces, (\ref{crv4m})--(\ref{crv0m}), give the background
geometry for the $\Sp(2n)$ or $\SU(k)$ curves with 4, 3, 2, 1, and 0
fundamentals, respectively.  Note that all surfaces (for generic $z$)
are non-degenerate tori.  Furthermore, by the arguments of sections
5.2 and 5.3, these curves should be equivalent to those of the
$\SU(2)$ theory with 4, 3, 2, 1, and 0 fundamentals, respectively.
For the lower numbers of flavors this is easy to check simply by
matching discriminants.

The 5 brane curves with decoupled fundamentals are found by taking the
same scaling limits as above.  For example, the reader can easily
check that the $\SU(2n)$ curve with equal mass $m_A$ antisymmetrics
(\ie\ no shift parameter) and no fundamentals is
\be\label{crv2n0f} 
0=\sum_{a=0}^n r_{2a} z^{n-a} +  
\left({y\over \hat x \hat w} - 2m_A \right)
\sum_{a=1}^n r_{2a-1} z^{n-a} 
\ee 
with $r_a$'s given by (\ref{radefn}), as found by taking the 
appropriate shifts, rescalings, and limits of (\ref{crv2n}).

\section{Deriving S duality for the N=4 SU(n) theory} 
 
As an application of the curves we have found, we will use them in
this section to derive part of the S duality identifications of the
$N=4$ supersymmetric $\SU(n)$ theory following the argument of
\cite{a9706}.  The basic idea is to realize the $N=4$ theory as an
infrared fixed point in an asymptotically free $N=2$ theory found by
appropriately tuning parameters and vevs in the asymptotically free
theory.  Then by scaling to the infrared the full scale-invariant
$N=4$ theory can be recovered.  In particular we can find the geometry
of the coupling space of the embedded $N=4$ theory from the Coulomb
branch of the asymptotically free $N=2$ theory.  The complex geometry
of the space of couplings is precisely the information encoded in the
S duality group \cite{a9706}.

The particular embedding we will use is the following.  Start with the
asymptotically free $N=2$ theory with gauge group $\SU(2n)$, 2
massless antisymmetric hypermultiplets, and strong coupling scale
$\Lambda$.  By appropriately tuning vevs on the Coulomb branch we can
higgs this theory at a scale $\mu$ to $\SU(n)\times\SU(n)\times\U(1)$
with two massless bifundamental hypermultiplets and with equal
dimensionless couplings in the two $\SU(n)$ factors.  The $\U(1)$
factor decouples, leaving the scale-invariant $\SU(n)\times\SU(n)$
elliptic model of \cite{w9703}.  Finally, by turning on a vev for one
of the bifundamentals (on its Higgs branch) we higgs the theory to the
diagonal $\SU(n)$ with a massless adjoint hypermultiplet---the $N=4$
theory.

Classically we can break $\SU(2n) \to \SU(n) \times \SU(n) \times
\U(1)$ on the Coulomb branch by giving the adjoint scalar $\Phi$ a vev
of the form
\be\label{11star}
\Phi = \mu \pmatrix{\bo_n & 0\cr 0&-\bo_n\cr}
\ee
where $\bo_n$ is the $n\times n$ identity matrix and $\mu$ is the
scale of the vev.  Under this breaking each antisymmetric
hypermultiplet $(A,\til A)$ decomposes as
\be
A = \pmatrix{A_1 & B\cr -B & A_2\cr}
\ee
and similarly for $\til A$, where $A_1$ and $A_2$ are antisymmetrics
under the first and second $\SU(n)$ factors respectively, while $B$ is
a bifundamental.  Writing the $N=2$ action in $N=1$ superfield
notation, each antisymmetric enters the superpotential as
\bea
{\cal W}_A &=& \mbox{tr} (A\Phi \til A) + m_A \mbox{tr}(A\til A)
= \mbox{tr} (A [\Phi + m_A \bo_{2n}]\til A) \nonumber\\
&=& \mbox{tr} \pmatrix{A_1 & B\cr -B & A_2\cr}
\pmatrix{m_A+\mu & 0 \cr 0 & m_A-\mu \cr} 
\pmatrix{\til A_1 & \til B \cr -\til B & \til A_2\cr} \\
&=& (m_A+\mu)\mbox{tr}(A_1\til A_1) + (m_A-\mu)\mbox{tr}(A_2\til A_2)
- 2m_A \mbox{tr}(B\til B) .\nonumber
\eea
Thus, if we keep the antisymmetric in the $\SU(2n)$ theory massless we
see that the antisymmetrics in the $\SU(n)$ factors become massive
while the bifundamentals remain massless, and we indeed get the scale
invariant $\SU(n)\times\SU(n)$ elliptic model.  Furthermore, a
one-loop renormalization group matching tells us that the couplings,
$q_{e1}$, $q_{e2}$, of the two $\SU(n)$ factors are given by (at weak
coupling)
\be\label{1lpmat}
q_{e1}=q_{e2} \sim {\Lambda^2 \over \mu^2}
\ee
and the equality between the two $q$'s is enforced by the symmetry
between the two $\SU(n)$ factors.  (The proportionality factor can be
found from a detailed one loop matching; but for the purposes of this
argument this factor is unimportant.)

Finally, upon giving one of the bifundamentals a vev we break
the $\SU(n)\times\SU(n)$ theory to the diagonal $\SU(n)$ and
the remaining bifundamental decomposes as a massless adjoint
hypermultiplet and a singlet (which is therefore decoupled).
This is precisely the $N=4$ $\SU(n)$ theory.  By a tree-level matching
its coupling $q$ is just
\be
q = q_{e1}q_{e2} \sim {\Lambda^4\over \mu^4} .
\ee

The curve describing the Coulomb branch of the $\SU(2n)$ theory with
two masseless antisymmetrics is given by (\ref{crv0m}),
(\ref{crv2n0f}), and (\ref{radefn}) with $m_A=0$.  In the $\hat x=1$
$\CP^2_{(1,1,2)}$ coordinate patch, and renaming $\hat w \to x$, these
curves read
\bea\label{crv4}
y^2 &=& -2x(x^2+\textstyle{1\over2}zx-\Lambda^4) ,
\nonumber\\
0 &=& \sum_{a=0}^n r_{2a} z^{n-a} +  
{y\over x} \sum_{a=1}^n r_{2a-1} z^{n-a} ,
\eea
where
\be
r_a = (-)^a \sum_{\ell_1 < \cdots < \ell_a}  
\phi_{\ell_1}\cdots\phi_{\ell_a} , \qquad 
r_0 = 1 , \qquad \sum_{\ell=1}^{2n} \phi_\ell \ = \ 0 .
\ee
To implement the higgsing (\ref{11star}) we should take in the curve
\be
\phi_\ell = \left\{
\begin{array}{ll}
+\mu+\varphi_{1,\ell} & \qquad \ell=1,\ldots,n \\
-\mu+\varphi_{2,(\ell-n)} & \qquad \ell=n+1,\ldots,2n 
\end{array}
\right.  \ee 
with $|\varphi_{i\ell}|\ll|\mu|$ and $\sum_{\ell=1}^n 
\varphi_{i\ell} = 0$.  Classically the $\varphi_{i\ell}$ are the
eigenvalues of the adjoint scalars for each $\SU(n)$ factor, and the
scale invariant theory is the ``origin'' of their Coulomb branch:
$\varphi_{i\ell}=0$ for $i=1,2$ and $\ell=1,\ldots,n$.  In fact, this
is true quantum mechanically as well.  For if we set
$\varphi_{i\ell}=0$ in the curve (\ref{crv4}), then it follows that
$r_{2a-1}=0$ and $r_{2a} = {n\choose a} (-\mu^2)^a$, so the 5 brane
curve becomes simply
\be
0 = (z-\mu^2)^n
\ee
which has the singularity expected of a scale invariant theory at
$z=\mu^2$.

We have therefore shown that the space of scale invariant
$\SU(n)\times \SU(n)$ elliptic models with equal couplings in the two
factors is described by a one-complex dimensional subspace of the
Coulomb branch of the $\SU(2n)$ theory given by (\ref{11star}).
Finally, by the breaking on the Higgs branch of one of the
bifundamentals we obtain the $N=4$ theory; furthermore, by an $N=2$
non-remormalization theorem \cite{aps9603} the Coulomb branch geometry
can not depend on hypermultiplet vevs, so the $\mu$ plane remains an
image of the space of couplings of the $N=4$ theory.

Actually, we have seen that at weak coupling the $N=4$ coupling $q
\sim \mu^{-4}$, so at weak coupling at least, the $\mu$-plane is in
fact a four-fold cover of the coupling space.  However, the
asymptotically free $\SU(2n)$ with two antisymmetrics theory has a
classical $\U(1)_R$ $R$-symmetry which is broken by instantons down to
a non-anomalous $\Z_4$ symmetry.  This $\Z_4$ global symmetry acts on
$\Phi$ by a phase rotation $\Phi \to e^{i\pi/2} \Phi$, and therefore
acts on $\mu$ in the same way.  This means that the coupling space is
really the $\mu$-plane modded out by this $\Z_4$ action, or,
equivalently, the $\mu^4$-plane instead, matching the weak coupling
result (\ref{1lpmat}).

By looking at the effective theory on the $\mu^4$ subspace we can
explore some features of the space of $N=4$ theories.  The curve
describing the Coulomb branch geometry is just (\ref{crv4}) at
$z=\mu^2$.  By rescaling $x\to \mu^2 x$ and $y \to \mu^3 y$ the
curve becomes the torus
\be\label{crv5}
y^2 = -2x \left( x^2 + {1\over2} x - {\Lambda^4\over \mu^4} \right) .
\ee
This is regular everywhere except at three points where the torus
degenerates: $\mu^4 = \infty$, $-16 \Lambda^4$, and $0$.  The $\mu^4 =
\infty$ point is just the weak coupling singularity.  The other two
are apparently ``infinitely strongly'' coupled singularities.  This
result shows that there must be S duality identifications on the naive
classical coupling space of the $N=4$ theory, since classically the
$N=4$ theory has a whole line of infinitely strongly coupled
singularities (Im$\tau = 0$ or $|q|=1$).  It should be emphasized
that these S duality identifications thus derived are exact
equivalences of the whole $N=4$ quantum theory, and not just of
low-energy or supersymmetric quantities, since they were derived from
a definition of the scale invariant theory as an infrared fixed point
of an asymptotically free theory \cite{a9706}.

Note, however, that the above derived S duality identifications need
not be the complete set of S dualities of the $N=4$ theory
\cite{a9706,ab9910}.  Indeed, the $N=4$ theory is believed to have an
exact $SL(2,\Z)$ group of identifications of its coupling $\tau$.  The
resulting coupling space---a fundamental domain of $\SL(2,\C)$---has
only one singularity which corresponds to weak coupling.  It is easy
to see that these further S duality identifications are consistent
with our curve.  Indeed, since the complex structure of any torus, and
our low-energy torus (\ref{crv5}) in particular, is invariant under
$\SL(2,\C)$ transformations, there will be a further set of
identifications of the coupling parameter $\Lambda^4/\mu^4$ which
identify the two ultra-strong coupling points with the weak coupling
point.  From the point of view of the specific asymptotically free
$N=2$ theory in which we have embedded the $N=4$ theory, these extra
identifications just look like accidental equivalences of the low
energy effective actions, and are not necessarily exact equivalences
enforced by a global (microscopic) symmetry.  It would be interesting
to see whether there are other embeddings of the $\SU(n)$ $N=4$ theory
which, together with the embedding shown here, allow one to derive the
full S duality group of the $N=4$ theory.

\acknowledgments It is a pleasure to thank Z. Kakushadze, J. Mannix,
K. Narayan, V. Sahakian and H. Tye for helpful comments and
discussions.  This work was supported in part by DOE grant
DOE-FG02-84ER40153, NSF grant PHY95-13717, by NSERC of Canada and
by FCAR of Quebec.

\end{document}